\documentstyle[12pt,epsfig,color]{article}

\def\tr{{\rm tr}}
\def\Tr{{\rm Tr}}

\def\I{{\cal I}}

\def\R{{\bf R}}
\def\C{{\bf C}}
\def\Z{{\bf Z}}
\def\N{{\bf N}}
\def\C#1{C^{(#1)}}
\def\half{\frac{1}{2}}
\def\RR{R-R }
\def\NSNS{NS-NS }
\def\RRT{R-R,T }
\def\NSNST{NS-NS,T }

\def\spc{\;\;\;\;}

\def\dh{$\hat{\mbox{D}}$}
\def\db{$\bar{\mbox{D}}$}

\newcommand{\rmd}{{\rm d}}
\def\tq{{\tilde q}}

\def\pmb#1{\setbox0=\hbox{#1}
 \kern-.025em\copy0\kern-\wd0
 \kern.05em\copy0\kern-\wd0
 \kern-.025em\raise.0433em\box0 }

\def\dh{$\hat{\mbox{D}}$}
\def\db{$\bar{\mbox{D}}$}
\def\3{\ss}
\def\sq{\hbox{\rlap{$\sqcap$}$\sqcup$}}
\def\qed{\ifmmode\sq\else{\unskip\nobreak\hfil
\penalty50\hskip1em\null\nobreak\hfil\sq
\parfillskip=0pt\finalhyphendemerits=0\endgraf}\fi}

\newcommand{\ket}[1]{|#1\rangle}
\newcommand{\bra}[1]{\langle#1|}

\def\mm {\advance\voffset by -1.5cm
\advance\hoffset by -2.1cm
\textwidth=17.3cm
\textheight=20.0cm}

\newcommand{\ms}[1]{\mbox{\scriptsize #1}}

\def\b,#1,#2(#3){\left|B#1,#2\right>_{#3}}

\def\gso#1,#2{\frac{1}{4}(1#1(-1)^F)(1#2(-1)^{{\tilde F}})}

\def\xxx#1           {{\sf hep-th/#1} }

\def\npb#1(#2)#3     {Nucl. Phys. {\bf B#1} (#2) #3 }
\def\rep#1(#2)#3     {Phys. Rept.{\bf #1} (#2) #3 }
\def\pla#1(#2)#3     {Phys. Lett. {\bf #1A} (#2) #3 }   
\def\plb#1(#2)#3     {Phys. Lett. {\bf #1B} (#2) #3 }
\def\prl#1(#2)#3     {Phys. Rev. Lett.{\bf #1} (#2) #3 }
\def\prd#1(#2)#3     {Phys. Rev. {\bf D#1} (#2) #3 }
\def\ap#1(#2)#3      {Ann. Phys. {\bf #1} (#2) #3 }
\def\rmp#1(#2)#3     {Rev. Mod. Phys. {\bf #1} (#2) #3 }
\def\cmp#1(#2)#3     {Comm. Math. Phys. {\bf #1} (#2) #3 }
\def\mpla#1(#2)#3    {Mod. Phys. Lett. {\bf A#1} (#2) #3 }
\def\ijmp#1(#2)#3    {Int. J. Mod. Phys. {\bf A#1} (#2) #3 }
\def\cqg#1(#2)#3     {Class. Quant. Grav. {\bf #1} (#2) #3 }
\def\am#1(#2)#3      {Adv. Math. {\bf #1} (#2) #3 }
\def\im#1(#2)#3      {Invent. Math. {\bf #1} (#2) #3 }
\def\jhep#1(#2)#3    {JHEP {\bf #1}(#2) #3 }
\def\npps#1(#2)#3    {Nucl.Phys.Proc.Suppl. {\bf #1}(#2) #3 }
\def\jgp#1(#2)#3     {J. Geom. Phys. {\bf #1}(#2) #3 }

\def\dstyler#1       {\documentstyle{report}[#1]}
\def\dstylea#1       {\documentstyle{article}[#1]}
\def\bd              {
\begin{document}}
\def\ed              {\end{document}}
\def\be	             {\begin{equation}}
\def\ee              {\end{equation}}
\def\ba              {\begin{eqnarray}}
\def\ea              {\end{eqnarray}}
\def\ni              {\noindent}
\def\bb#1            {\begin{thebibliography}{#1}}
\def\eb              {\end{thebibliography}}


\def\etal {{\em et al.} }
\def\w    {\;_\wedge}
\def\ie   {{\em i.e.}}
\def\ibid {{\em ibid.}}
\def\cf   {{\em c.f.} }

\mm

\bd

\thispagestyle{empty}
\def\thefootnote{\fnsymbol{footnote}}
\begin{flushright}
  hep-th/0110041 \\
  AEI-2001-115
\end{flushright}

\vskip 0.5cm

\begin{center}\LARGE
{\bf Dirichlet Branes on Orientifolds}
\end{center}

\vskip 1.0cm

\begin{center}
{\large N. Quiroz\footnote{E-mail address: {\tt quiroz@aei.mpg.de}}
\footnote{On leave from Departamento De F\'isica, Centro de Investigaci\'on y Estudios 
Avanzados del IPN (CINVESTAV-IPN), M\'exico D.F., M\'exico}, 
B. Stefa\'nski, jr.\footnote{E-mail address: {\tt bogdan@aei.mpg.de}}}

\vskip 0.5cm

{\it Max-Planck-Institut f\"ur Gravitationsphysik, Albert-Einstein 
Institut \\ Am M\"uhlenberg 1, D-14476 Golm, Germany}
\end{center}

\vskip 1.0cm

\begin{center}
October 2001
\end{center}

\vskip 1.0cm

\begin{abstract}

\noindent We consider the classification of BPS and non-BPS D-branes in orientifold models. 
In particular we construct all stable BPS and non-BPS D-branes in the Gimon-Polchinski (GP)
and Dabholkar-Park-Blum-Zaffaroni (DPBZ) orientifolds and determine  their stability regions 
in moduli space as well as decay products.
We find several kinds of integrally and torsion 
charged non-BPS D-branes. Certain 
of these are found to have projective representations of the 
orientifold~$\times$~GSO group on the Chan-Paton factors. 
It is found that the GP orientifold is not described by equivariant 
orthogonal K-theory as may have been at first expected. Instead a twisted 
version of this K-theory is expected to be relevant.
\end{abstract}

\vfill

\setcounter{footnote}{0}
\def\thefootnote{\arabic{footnote}}
\newpage

\renewcommand{\theequation}{\thesection.\arabic{equation}}

\section{Introduction and Summary}\label{sec1}
\setcounter{equation}{0}

The classification of D-branes is an important aspect of improving our understanding 
of string theory. It gives a better handle on what vacua are 
allowed, as well as on how these can decay into one another via tachyon condensation. 
For each model a suitable K-theory should exist and constructing branes from 
physical principles allows to test the K-theory hypothesis more extensively. Furthermore, minimally 
charged, stable non-BPS D-branes should
have descriptions in dual theories and finding these is an important step in testing dualities 
beyond the BPS level. Finally, D-branes provide a way of introducing 
non-abelian gauge fields as well as a way of breaking supersymmetry, giving them phenomenological
relevance. In this paper we investigate D-branes in orientifold theories. We find
a surprisingly rich spectrum of BPS and non-BPS D-branes and discuss decays and transitions 
between them.  We find large classes of torsion charged D-branes. It is precisely for 
torsion charges where K-theory and rational cohomology may differ allowing 
for non-trivial tests of K-theory.

Since 1995 Dirichlet branes~\cite{PolRR} have played a central role in string theory. 
Initially D-branes were constructed as BPS objects with corresponding supergravity solutions. 
These were instrumental in testing dualities at the BPS level (for a review see~\cite{Polrev}).
Later it was realised that, in the first-quantised theory, stable non-BPS D-branes 
too could be constructed as coherent states in the closed string theory~\cite{Snbps,BG,WittK,Frau,GS}. 
Such non-BPS D-branes have provided several non-BPS tests of dualities~\cite{BGhet,DS,GabSak}. 
The description of D-branes by boundary states in a conformal field 
theory~\cite{CLNY-1,PC,CLNY,sewing} has been particularly useful when a geometric picture is not 
apparent~\cite{nongeom}.

Non-BPS D-branes often decay into BPS-anti-BPS pairs of D-branes when certain closed string
marginal deformations are turned on. These brane descent relations
have been shown to be marginal deformations in the open string
conformal field theory~\cite{Snbps} and as such conserve mass and charge in the process.
The decays in the first quantised theory 
suggested that D-branes can annihilate with anti-D-branes into the closed string vacuum~\cite{Snbps,sred} 
via tachyon condensation.
This led to the description of D-branes in terms of K-theory~\cite{WittK,MM} where such 
annihilations are a central feature. A classification of D-branes via K-theory guarantees
Dirac quantisation of charges. It may also provide a description of torsion charged D-branes
where it is not always obvious how to characterise the charges (and decay products) of 
various (unstable) brane configurations.  For torsion charges K-theory in general differs from 
cohomology. As such torsion D-branes offer a confirmation of the K-theory 
perspective. In the orbifolds of oriented theories studied so far no 
torsion branes were found. On the other hand the simplest orientifold
- Type I theory - has four such branes the D-1,0,7- and 8-branes. 

In this paper we classify D-branes in the 
$\Omega\times\I_4$ orientifold of Type IIB theory first considered by~\cite{BS}. 
More recently these models were extensively investigated in the D-brane language by
Gimon and Polchinski~\cite{GP} (see also~\cite{Many}) and Dabholkar, Park and Blum, 
Zaffaroni~\cite{DP,BZ}. Both models can be regarded either as orbifolds of Type I or as $\Omega$ 
projections of the $\I_4$ orbifold of Type IIB. Projecting by $\Omega$ 
does not introduce any new closed 
string states while there are $\I_4$-twisted closed string sectors to 
which the D-brane may couple. Since we will construct D-branes as 
boundary states from various closed string sectors it is particularly 
useful to think of the GP orientifold as the $\Omega$-projected K3 
orbifold of Type IIB on $\R^6\times T^4$. D-branes in the orientifold are $\Omega$-invariant 
configurations of D-branes in the orbifold. Let us briefly review the kinds 
of D-branes present in the orbifold and discuss the action of $\Omega$ on 
them. In the orbifold~\cite{GS} it is convenient to
label D$p$-branes as $(r,s)$-branes where $s$, $r+1$  are the number of 
directions along which 
the brane extends on which $\I_4$ does or does not act, respectively, 
with $r+s=p$. There are two elementary types of integrally charged\footnote{There are no stable
torsion branes in the K3 orbifold.} D-branes in this IIB orbifold: 
for $s$ even there are BPS fractional 
branes while for $s$ odd there are non-BPS truncated branes; in both 
cases $r$ is odd.
The fractional branes are charged under both untwisted and twisted \RR 
charges coming from
anti-symmetric forms
\be
C^{(r+1,s)}\,,\qquad C^{(r+1)}_t\,,
\ee
while the truncated branes are charged only under the twisted \RR forms. Here
$C^{(r+1,s)}$ are the usual Type IIB \RR $r+s+1$-forms with $r,s$ indices 
in the non-compact and internal directions, respectively; $C^{(r+1)}_t$ 
are the twisted \RR $r+1$-forms which have indices in the non-compact 
directions only. An $(r,s)$-brane couples to $2^s$ such twisted sectors. 
The BPS branes are stable - they do not develop open string tachyons on 
their world-volume. On the other hand the non-BPS branes have open string 
winding or momentum tachyons for certain values of the compactification 
radii and decay into BPS-anti-BPS pairs of fractional 
branes. Note also that two fractional branes with the same untwisted \RR 
charge and opposite twisted \RR charges can come together and move of the 
fixed points to form a bulk BPS brane charged only under the untwisted 
\RR sector.

$\Omega$ keeps only certain of the above \RR forms. In particular in the untwisted 
sector  
\be
C^{(r+1,s)}\,,\qquad\mbox{with } r+s=1,5,9
\ee
survive. In the twisted sector two choices are allowed; either we keep a 
hypermultiplet at each fixed point as in~\cite{GP} \ie
\be
C^{(r+1)}\,,\qquad\mbox{with } r=-1,3
\label{rrtwom}
\ee
survive or we keep a tensor multiplet~\cite{DP,BZ} in which case
\be
C^{(r+1)}\,,\qquad\mbox{with } r=1,5\,,
\ee
are $\Omega$ invariant.
For definiteness we concentrate on the GP orientifold throughout this 
paper - in section~\ref{sec7} we discuss for completeness the DPBZ orientifold.

D-branes are stable due to the charges that they carry, and since only  
$\Omega$-even configurations of branes from the orbifold are allowed in 
the orientifold, the spectrum of stable branes changes. Since truncated branes 
are charged only under twisted \RR fields the $\Omega$ projection will either keep or remove 
them.\footnote{Strictly speaking a brane may still 
be stable even though it carries no integral charge - there may be some 
torsion charge which stabilises it. For now we concentrate on integrally 
charged branes and discuss torsion charges below.} In particular 
from~(\ref{rrtwom}) we see that the truncated $(r,s)$-branes with 
$r=-1,3$ are present in the orientifold while those with $r=1,5$ are 
removed.\footnote{$s$ is odd since these are truncated branes in the 
orbifold~\cite{GS}.}

Fractional branes carry twisted and untwisted \RR charges and $\Omega$ 
can be odd or even on these separately. As a result there are four 
possible destinies for the
fractional branes of the orbifold in the orientifold:
\begin{itemize}
\item $r=-1,3$, $s=0,4$: the bulk \RR charge is not $\Omega$ invariant - 
the corresponding branes are only charged under twisted \RR charges, are 
non-BPS and very similar to the usual truncated branes;
\item $r=-1,3$, $s=2$: these fractional branes are $\Omega$ invariant;
\item $r=1,5$, $s=0,4$: the twisted charge is projected out - the branes 
are BPS but do not carry any twisted charges and can be thought of as a 
pair of fractional branes in the orbifold with opposite twisted charges. 
As we will see later such a pair of branes is nonetheless not allowed to 
move off the fixed points and we will refer to them as {\em stuck}. Note that 
for $r=5$ and $s=0,4$ these are precisely the tadpole canceling 5- and 
9-branes of the GP model;
\item $r=1,5$, $s=2$: both twisted and untwisted \RR charges are 
projected out by $\Omega$ - there are no integrally charged branes.
\end{itemize}

The lack of integral charges does not, in general, mean that a brane is 
unstable. There may be some torsion charge stabilising the brane as 
in the case of the D0- or D7-brane of Type I.
The former is stable in the tadpole cancelled theory, while the latter is 
unstable due to the tachyonic open strings that stretch between it and 
the tadpole canceling D9-branes. The D7-brane decays into a gauge 
configuration on the D9-brane world-volume~\cite{flux}, demonstrating that even though 
the brane is unstable the corresponding charge is present in the theory - 
after all K-theory 
only predicts the presence of a charge rather than determine what kind of 
object carries it.

In the GP orientifold we encounter a large number of torsion charged branes. 
For example, the simplest $\Omega$ invariant pair of fractional 
$(5,2)$-branes in the Type II orbifold, couples to the twisted and untwisted \NSNS sectors,
 has opposite twisted and untwisted 
\RR charges and decays into the vacuum in the orbifold. Hence  above we 
were lead to conclude that there are no integrally charged $(5,2)$-branes 
in the orientifold. In fact in the orientifold the configuration
does carry a torsion charge and as a 
result is stable (in other words there are no tachyonic open strings that 
have both end-points on the (5,2)-brane configuration). 
In the non-compact theory the charge is $\Z_2\oplus\Z_2$.
As in the D7-brane of Type I the open strings 
that stretch between the tadpole canceling D9-branes and the (5,2)-brane 
have tachyons in their spectrum. We expect that as a result the 
(5,2)-brane does decay. However, this semi-stability of the (5,2)-brane 
indicates the presence of a torsion charge in the GP orientifold. In the 
DPBZ orientifold similar torsion branes are in fact stable in the tadpole cancelled theory.  
We also find a number of $\Z_2$ charged D-branes. These have $r=5$ and $s=1,3$. 
We will show that there are decay channels between these indicating that 
they carry the same torsion charge.

Type II orbifolds are classified by equivariant complex K-theory. In all 
of the cases considered K-groups computed did indeed agree with the 
D-brane spectrum. Type I D-branes on the other hand are classified by 
orthogonal K-theory, which is also in agreement with the D-branes 
present in the theory.\footnote{The D7 and D8-branes are somewhat subtle, 
and we refer the reader to~\cite{flux} for discussions of these.} One might 
expect that orbifolds of Type I should be classified by orthogonal 
equivariant K-theory. From the string theory perspective though it is 
clear that there is a potential subtlety in defining the action of 
$\Omega$ on the closed string twisted sectors. As we have already seen in 
the $\I_4\times\Omega$ orientifolds there are two ways of 
defining the action of $\Omega$ on the $\I_4$-twisted sectors, which 
leaves either a tensor- or a hypermultiplet in the orientifold. This is 
very similar to the discrete torsion~\cite{vafadt} one encounters in $\Z_2\times\Z_2$ 
Calabi-Yau orbifolds~\cite{z2torsion}. In the Calabi-Yau models there are two theories 
with $(h^{11},h^{21})=(3,51)$ and $(51,3)$ which arise due to the 
presence of projective representations of the orbifold group 
$\Z_2\times\Z_2$. The models are described by the equivariant complex 
K-theory and its twisted version. In the orientifold models considered in 
this paper, the orientifold group is also $\Z_2\times\Z_2$, and indeed we 
have two possible models as in the Calabi-Yau orbifolds. As a result the 
equivariant orthogonal K-theory can only describe one of these. We 
show that it in fact gives the DPBZ model. The second model should be 
described by a suitably twisted version of this K-group, which 
to the best of our knowledge has not been studied in the mathematical literature.

The paper is organised as follows. In section~\ref{sec2} we discuss the 
construction of GSO, orbifold and $\Omega$ invariant boundary states in 
each of the closed string sectors; from these we construct boundary 
states corresponding to BPS D-branes in the orientifold: fractional, 
stuck and bulk. In sections~\ref{sec3} and~\ref{sec4} boundary states 
corresponding to integrally and torsion charged non-BPS D-branes are 
constructed. The stability conditions and decay channels of 
the branes are also analysed. In section~\ref{sec5} the open string 
perspective is discussed. In particular open strings that end on 
many of the branes constructed in the closed string channel have 
non-trivial Chan-Paton factors; the representation of the orientifold 
group as well as the GSO projection on these is discussed. For example 
$(-1)^F$ and $\Omega$ are found to form a projective representation on 
the Chan-Paton factors of the D7-brane of Type I. We show how this phase is 
compensated for by an opposite one on the world-sheet fields to give a 
genuine representation on the full open string Hilbert space. 
Section~\ref{sec6} contains a discussion of the relevant K-theories for 
the GP and DPBZ models. In particular it is argued that the equivariant 
orthogonal K-theory $KO_{\Z_2}$ is relevant to DPBZ rather than GP, where 
a twisted version of $KO_{\Z_2}$ has to be defined.
In Section~\ref{sec7} branes on DPBZ are discussed and section~\ref{sec8} 
contains some conclusions and open problems. We have included several 
appendices which contain computational details to which we refer to 
throughout the paper. 

In~\cite{EP} certain aspects of D-branes in the GP model have been 
studied. In particular orientifold invariant boundary states in each of 
the closed string sectors were constructed. Further, the truncated branes 
were identified, though the stability of these branes differs from 
our results. We have also found a different, and much bigger, set of 
torsion charged branes to~\cite{EP}. 

\section{BPS D-branes}\label{sec2}
\setcounter{equation}{0}

D-branes interact with closed string states, and as such can be described 
as coherent states in the closed string Fock space~\cite{CLNY-1,PC,CLNY}. Since these represent 
a boundary in the world-sheet at which closed strings are absorbed or 
emitted, these coherent states are called boundary states. In each of these 
closed string sectors one constructs a GSO invariant coherent state. In 
orbifolds and orientifolds these boundary states also have 
to be invariant under the action of the symmetry group being 
modded out. Hence only GSO- and orbifold/orientifold- 
invariant closed string states couple to D-branes. These invariances 
place restrictions on the allowed boundary states. D-branes are also 
hypersurfaces on which open strings end; only certain combinations of boundary states 
give rise to a consistent open string spectrum. In this section we discuss the construction of 
orientifold invariant boundary states, and combine boundary states from 
various sectors to construct BPS branes. 

\subsection{Orientifold invariant boundary states}\label{sec21}

In each closed string sector one constructs two boundary states, 
corresponding to the two spin structure choices for fermions on the 
boundary. GSO invariance ensures that at most one linear combination 
survives. Orbifold and $\Omega$ invariance places further restrictions on 
the allowed $r$ and $s$ values. GSO and $\I_4$ invariance have been 
discussed in detail before~\cite{GS}, and in~\cite{EP} $\Omega$ 
invariance has been investigated. In appendix~\ref{appb} we review 
$\Omega$ invariance and summarise the results here. 
In the twisted sectors $\Omega$ can act in one of two ways, giving the GP 
or DPBZ orientifolds; here we state the results for the GP orientifold. Boundary 
states in each closed string sector are $\Omega$, $I_4$ and GSO invariant for
\begin{itemize}
\item $\ket{B(r,s)}_{\mbox{\scriptsize\NSNS}}$ all $(r,s)\,,$
\item $\ket{B(r,s)}_{\mbox{\scriptsize\RR}}$ $r+s=1,5,9\,,$ 
\item $\ket{B(r,s)}_{\mbox{\scriptsize\NSNS,T}}$ $s=2\,,$ 
\item $\ket{B(r,s)}_{\mbox{\scriptsize\RR,T}}$ $r=-1,3\,.$
\end{itemize}
A D-brane is a consistent linear combination of boundary states from 
various closed string sectors. The consistency conditions determine the 
normalisation of boundary states as well as the allowed linear combinations 
from different sectors.  In an oriented theory these consistency conditions 
come from the requirement that the cylinder diagram reproduce the annulus partition 
function of an open string ending on the D-brane with a projection 
operator inserted into the trace. For example a D-brane coupling to 
the untwisted and twisted \RR sectors would not lead to a consistent open 
string partition function, as the operator inserted in the corresponding 
open strnig trace would be proportional to $(-1)^F+\I_4(-1)^F$ which is not a projection operator; a 
brane that couples to both these sectors needs to couple to the untwisted 
and twisted \NSNS sectors as well.

In theories with $\Omega$ projections there are orientifold planes 
represented by crosscaps (for a construction of these coherent states see 
Appendix~\ref{appa}); for example in the GP orientifold there are O9- and 
O5-planes. A coherent state corresponding to a consistent D-brane now has 
to reproduce the annulus and M\"obius strip contributions to the one-loop 
partition function for an open, unoriented string ending on the D-brane. 
The annulus comes from the cylinder diagram corresponding to the exchange 
of a closed string between two boundary states, while the M\"obius strip 
comes from the exchange of a closed string state between a crosscap and a 
boundary state.
In practice the square of the normalisation of the D-brane and O-plane is 
obtained from the annulus and Klein bottle diagrams, while the relative 
sign follows from the M\"obius strip. In~\cite{GS} the normalisations of 
various boundary states were worked out rather explicitly. The 
computations are easily generalised to the case studied presently as 
summarised in Appendix~\ref{appc}.

\subsection{BPS D-branes in the GP orientifold}

There are two kinds of elementary BPS D-branes in the GP orientifold - 
the fractional and stuck branes. Both are located at the transverse fixed 
points and can only have discrete Wilson lines in the compact directions 
along which they extend. They differ in that the fractional branes 
couples to both twisted and untwisted closed string sectors while the 
stuck branes couple only to the untwisted sectors. In this subsection we 
discuss both kinds of branes and comment on how bulk BPS branes come into 
the picture.

Fractional branes are charged under both twisted and untwisted \RR 
charges and so, as mentioned at the end of the previous subsection, in 
order that the open string partition function have a consistent 
projection operator inserted, these branes couple to all closed string 
sectors. In the GP orientifold there are only two such branes:
the $(r,s)=(-1,2)$ and $(3,2)$, described by the boundary states
\ba
\ket{D(r,s)} & = & {\cal N}_{(r,s),U}
\left(\left|B(r,s)\right>_{\mbox{\scriptsize\NSNS}} 
+ \epsilon_1 \left|B(r,s)\right>_{\mbox{\scriptsize\RR}} \right) 
\nonumber \\
& & \quad 
+ \epsilon_2 \; {\cal N}_{(r,s),T} \sum_{\alpha=1}^{2^s}
e^{i\theta_\alpha}
\left(\left|B(r,s)\right>_{\mbox{\scriptsize\NSNS,T$_\alpha$}} 
+ \epsilon_1 
\left|B(r,s)\right>_{\mbox{\scriptsize\RR,T$_\alpha$}} \right)\,,
\label{fractboun}
\ea
where $\epsilon_i=\pm 1$, $\alpha$ labels the different fixed points 
between which the
brane stretches, and $\theta_\alpha=0,\pi$ is the Wilson
line that is associated to the difference of the fixed point $\alpha$
and the origin. ${\cal N}_{(r,s),U}$ and ${\cal N}_{(r,s),T}$ are the 
normalisations of the untwisted and twisted sectors, determined 
most easily by requiring that the closed string cylinder diagram 
reproduce the one-loop open string partition function
\be
\int_0^\infty dl\bra{D(r,s)}e^{-l\pi H_c}\ket{D(r,s)}
=\int\frac{dt}{2t}\Tr_{\mbox{\scriptsize 
NS-R}}(\frac{1}{2}\frac{1+(-1)^F}{2}\frac{1+\I_4}{2}
e^{-2\pi t H_o})\,,
\ee
where $t=1/2l$, $H_{c,o}$ are the closed and open string Hamiltonians and 
the extra factor of $1/2$ comes from the $\Omega$ 
projection.\footnote{The full open string partition function has three 
projections - GSO, orbifold and $\Omega$; the cylinder diagram is 
proportional to $1$ and the M\"obius strip to $\Omega$.} As a result the 
normalistions naively would differ from~\cite{GS} by an extra factor of 
$1/\sqrt{2}$, as is indeed the case for the $(-1,2)$-brane (see Table~\ref{tb1}). The $(3,2)$-brane 
though is a 5-brane and for it the smallest representation of $\Omega$ on Chan-Paton 
factors is the $2\times 2$ anti-symmetric matrix
\be
\gamma_\Omega=\left(\begin{array}{cc}0&i\\-i&0\end{array}\right)\,.
\ee
This follows~\cite{GP} from the fact that for D5-branes on world-sheet 
fields $\Omega^2=-1$ which is compensated for by the same relation on the 
Chan-Paton factors with the above matrix; on the full open string 
Hilbert space $\Omega$ squares to one. Similar phases on 
the sub-sectors of the open string Hilbert space are quite common and we 
discuss them in section~\ref{sec5}. Here we note that since the 
(3,2)-brane has $2\times 2$ Chan-Paton factors its normalisation is 
actually a $\sqrt{2}$ bigger than that of the $\I_4$-orbifold $(3,2)$-brane; this ensures 
that the Dirac quantisation condition is minimally satisfied. The 
situation is rather reminiscent of the D1-D5 discussion in Type I. Further, it is 
straightforward to check that in the open string world-volume theories of these fractional 
branes there are no
massless transverse scalars in the orbifolded directions and as a result 
the branes cannot move off the fixed-points.

A second class of BPS branes (which also cannot move off the fixed points) 
does not couple to twisted sectors.\footnote{This can be contrasted with 
orbifold theories where branes stuck at fixed points do couple to twisted 
sectors.} We refer to these branes as stuck branes. In the orbifold they 
are a pair of fractional branes with the same untwisted \RR 
charge, but opposite twisted \RR charges and are mapped to one another by 
$\Omega$, in other words  they form a bulk brane with a single transverse 
scalar in each of the compact 
directions; this scalar is removed by the $\Omega$ projection giving in the orientifold a 
BPS brane that couples only to the untwisted sectors, yet is stuck at the 
fixed points. From section~\ref{sec21} it is immediate that such branes 
exist for $r=1,5$ and $s=0,4$ and are described by the boundary state
\be
\ket{D(r,s)}  =  {\cal N}_{(r,s),U}
\left(\left|B(r,s)\right>_{\mbox{\scriptsize\NSNS}} 
+ \epsilon \left|B(r,s)\right>_{\mbox{\scriptsize\RR}} \right) \,,
\ee
with $\epsilon=\pm 1$ and the normalisation constant ${\cal N}_{(r,s),U}$ 
given in Appendix~\ref{appc}.\footnote{For branes with $r+s=5$ the 
Chan-Paton factors are $2\times 2$ as in the case of the $(3,2)$-brane.} 
In particular note that the GP orientifold has sixteen stuck $(5,0)$- and 
$(5,4)$-branes which cancel the untwisted 6- and 10-form \RR tadpoles. Further, neither
the O-planes nor the tadpole canceling stuck branes couple to the twisted \RR sectors. 
This is consistent with the lack of a twisted sector six-form tadpole in the GP 
model~\cite{GP,Many}.

Finally, we note that single bulk branes are simply pairs of fractional or stuck 
branes which do not couple to twisted closed string sectors, and carry untwisted \RR charges. 
It can be easily checked that such objects have a single transverse scalar in each 
of the compact directions which allows them to move off the fixed points.

\section{Integrally charged non-BPS \dh-branes}\label{sec3}
\setcounter{equation}{0}

Perturbatively stable D-branes have no open string tachyons on their 
world-volumes and carry an integral or torsion charge classified by a 
suitable K-theory. Integrally charged D-branes couple to suitable 
anti-symmetric \RR forms with which the charge is associated; in other 
words 
one may find an element of a rational cohomology from which the D-brane 
K-theory class was lifted. In the previous section we discussed integrally 
charged BPS branes in the GP model.\footnote{BPS D-branes have to be 
integrally charged since they are massive.}
In this section we construct the remaining stable integrally charged 
branes in the GP model. As in the $\I_4$ orbifold~\cite{Snbps,BG,GS} these are truncated 
non-BPS branes which couple to the untwisted \NSNS and twisted \RR 
sectors 
\be
\ket{\mbox{\dh}(r,s)} = {\cal 
N}_{(r,s),U}\left|B(r,s)\right>_{\mbox{\scriptsize\NSNS}} 
+ \epsilon \; {\cal N}_{(r,s),T} \sum_{\alpha=1}^{2^s}e^{i\theta_\alpha}
\left|B(r,s)\right>_{\mbox{\scriptsize\RR,T$_\alpha$}}\,,
\label{truncbrane}
\ee
and are only stable for certain values of the radii.\footnote{They are 
non-BPS as supersymmetry would require the presence of untwisted R-R 
(twisted NS-NS)  sector boundary state to combine with the untwisted 
NS-NS (twisted R-R) one. Further the one-loop partition function for open 
strings ending on the brane does not generically vanish.} Above ${\cal N}$ are the normalisations of 
the boundary states, $\epsilon=\pm 1$ and $\theta_\alpha=0,\pi$ are the Wilson lines 
associated to fixed point $\alpha$. The open strings 
that end on such branes have the projection
\be
\frac{1+\Omega}{2}\frac{1+(-1)^F\I_4}{2}\,.
\ee
The $r=-1,3$ and $s=1,3$ ($r=1,5$ and $s=1,3$) truncated branes of the $\I_4$ orbifold are 
(not) $\Omega$ invariant and hence are (not) present in the GP orientifold.  
The second type of truncated branes are an $\Omega$ invariant pair of 
fractional $r=-1,3$, $s=0,4$
branes from the $\I_4$ orbifold. For these values of $r$ and $s$ (see section~(\ref{sec21}))
The \RR untwisted and \NSNS twisted 
boundary states are $\Omega$ odd, so such pairs of fractional branes 
couple to the untwisted \NSNS and twisted \RR sectors only, justifying 
their name. 

Since both kinds of truncated branes are non-BPS their stability is not 
guaranteed. Indeed, while $(1+(-1)^F\I_4)/2$ removes the ground-state 
tachyon from the spectrum of the open string ending on a truncated brane, 
it keeps winding and momenta states of the form
\be
\ket{\frac{n}{R_{||}}}-\ket{-\frac{n}{R_{||}}}\,,\qquad\ket{wR_\perp}-\ket{-wR_\perp}\,,
\label{pottach}
\ee
where $n,w\in\Z$ and $R_{||},R_\perp$ are radii of the compact directions 
parallel, transverse to the D-brane, respectively. The lightest such 
states are non-tachyonic for
\be
R_{||}\le\sqrt{2}\,,\qquad R_\perp\ge\frac{1}{\sqrt{2}}\,,
\label{orbstab}
\ee
and outside of these regions the truncated branes of the orbifold decay 
into BPS-anti-BPS pairs of fractional branes~\cite{Snbps,BG,GS}. In the GP orientifold the 
open strings that end on truncated branes are projected by 
$(1+(-1)^F\I_4)/2$ as well as $(1+\Omega)/2$ and it is immediate that 
this second projection removes either the momentum or the winding states 
of~(\ref{pottach}) depending on the action of $\Omega$ on the open-string 
ground state.\footnote{In the above discussion we have taken the 
Chan-Paton factors to be trivial.} 
A good way to determine which states are removed is by computing the 
one-loop open string partition function in the closed string channel; 
since the computation is 
somewhat tedious we defer it to appendix~\ref{appc} and summarize the 
results below. Minimally charged $r=-1$, $s=0,1$ truncated branes are 
stable for all values of $R_{||}$ and
\be
R_\perp\ge\frac{1}{\sqrt{2}}\,,
\ee
while the $r=-1$, $s=3,4$ branes are stable for all values of $R_\perp$ 
and\footnote{Under four T-dualities in the internal directions  an 
$(-1,s)$-brane becomes a $(-1,4-s)$-brane and the stability conditions 
are T-duality invariant with a winding mode becomes a momentum mode.}
\be
R_{||}\le\sqrt{2}\,.
\ee
Non-minimally charged $r=-1$ branes have non-trivial CP factors and  
there will remain at least one winding and one momentum mode of the 
form~(\ref{pottach}). The results of Appendix~\ref{appc} confirm that 
these branes have the same stability as in the 
orbifold~(\ref{orbstab}).\footnote{Since truncated branes have 
non-vanishing forces between them the stability of a truncated brane with 
non-minimal \RR charge is somewhat complicated. By the stability of the 
non-minimally charged objects we simply mean the lack of open string tachyon in 
a hypothetical bound state.} As we will see below $r=3$ truncated branes 
have non-trivial Chan-Paton factors and so have the same stability 
conditions as the orbifold~(\ref{orbstab}).

We now turn to the discussion of the decay channels of a minimally 
charged $r=-1$ brane. As was mentioned above the $s=3,4$ branes are 
T-dual to the $s=1,0$-branes and so we discuss only the latter. 
In the discussion of decays throughout this paper we will 
compare the mass and charge (if any) of the decaying brane with that of 
the proposed decay products at the critical radius. If the two agree we 
shall take it to mean that the decay is via a marginal deformation in the 
conformal field theory corresponding to turning on a vev for a suitable 
tachyonic mode which becomes massless at the critical radius~\cite{Snbps}.

\begin{figure}[htb]
\begin{center}
\begin{picture}(0,0)%
\includegraphics{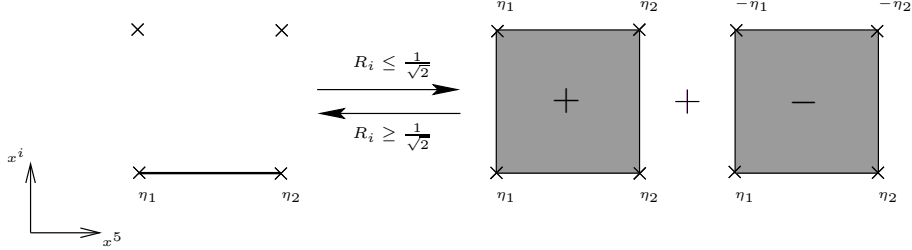}%
\end{picture}%
\setlength{\unitlength}{3947sp}%
\begingroup\makeatletter\ifx\SetFigFont\undefined%
\gdef\SetFigFont#1#2#3#4#5{%
  \reset@font\fontsize{#1}{#2pt}%
  \fontfamily{#3}\fontseries{#4}\fontshape{#5}%
  \selectfont}%
\fi\endgroup%
\begin{picture}(5522,1575)(1,-1486)
\put(601,-1486){\makebox(0,0)[lb]{\smash{\SetFigFont{5}{6.0}{\rmdefault}{\mddefault}{\updefault}{\color[rgb]{0,0,0}$x^5$}%
}}}
\put(  1,-961){\makebox(0,0)[lb]{\smash{\SetFigFont{5}{6.0}{\rmdefault}{\mddefault}{\updefault}{\color[rgb]{0,0,0}$x^i$}%
}}}
\put(3976,-1186){\makebox(0,0)[lb]{\smash{\SetFigFont{5}{6.0}{\rmdefault}{\mddefault}{\updefault}{\color[rgb]{0,0,0}$\eta_2$}%
}}}
\put(5476,-1186){\makebox(0,0)[lb]{\smash{\SetFigFont{5}{6.0}{\rmdefault}{\mddefault}{\updefault}{\color[rgb]{0,0,0}$\eta_2$}%
}}}
\put(4576, 14){\makebox(0,0)[lb]{\smash{\SetFigFont{5}{6.0}{\rmdefault}{\mddefault}{\updefault}{\color[rgb]{0,0,0}$-\eta_1$}%
}}}
\put(3076, 14){\makebox(0,0)[lb]{\smash{\SetFigFont{5}{6.0}{\rmdefault}{\mddefault}{\updefault}{\color[rgb]{0,0,0}$\eta_1$}%
}}}
\put(4576,-1186){\makebox(0,0)[lb]{\smash{\SetFigFont{5}{6.0}{\rmdefault}{\mddefault}{\updefault}{\color[rgb]{0,0,0}$\eta_1$}%
}}}
\put(826,-1186){\makebox(0,0)[lb]{\smash{\SetFigFont{5}{6.0}{\rmdefault}{\mddefault}{\updefault}{\color[rgb]{0,0,0}$\eta_1$}%
}}}
\put(1726,-1186){\makebox(0,0)[lb]{\smash{\SetFigFont{5}{6.0}{\rmdefault}{\mddefault}{\updefault}{\color[rgb]{0,0,0}$\eta_2$}%
}}}
\put(3076,-1186){\makebox(0,0)[lb]{\smash{\SetFigFont{5}{6.0}{\rmdefault}{\mddefault}{\updefault}{\color[rgb]{0,0,0}$\eta_1$}%
}}}
\put(2176,-811){\makebox(0,0)[lb]{\smash{\SetFigFont{6}{7.2}{\rmdefault}{\mddefault}{\updefault}{\color[rgb]{0,0,0}$R_i\ge\frac{1}{\sqrt{2}}$}%
}}}
\put(2176,-361){\makebox(0,0)[lb]{\smash{\SetFigFont{6}{7.2}{\rmdefault}{\mddefault}{\updefault}{\color[rgb]{0,0,0}$R_i\le\frac{1}{\sqrt{2}}$}%
}}}
\put(3976, 14){\makebox(0,0)[lb]{\smash{\SetFigFont{5}{6.0}{\rmdefault}{\mddefault}{\updefault}{\color[rgb]{0,0,0}$\eta_2$}%
}}}
\put(5476, 14){\makebox(0,0)[lb]{\smash{\SetFigFont{5}{6.0}{\rmdefault}{\mddefault}{\updefault}{\color[rgb]{0,0,0}$-\eta_2$}%
}}}
\put(4913,-646){\makebox(0,0)[lb]{\smash{\SetFigFont{14}{16.8}{\rmdefault}{\mddefault}{\updefault}{\color[rgb]{0,0,0}$-$}%
}}}
\put(3426,-629){\makebox(0,0)[lb]{\smash{\SetFigFont{14}{16.8}{\rmdefault}{\mddefault}{\updefault}{\color[rgb]{0,0,0}$+$}%
}}}
\end{picture}
\end{center}
\caption{{\scriptsize
The decay of a truncated (-1,1)-brane into a D\db-pair of (-1,2)-branes. $\eta_i=\pm 1$ are the
signs of the twisted \RR charges at the fixed points indicated by the crosses. The \dh$(-1,1)$-brane
has $\epsilon=\eta_1$ and $\theta_5=\pi\eta_2/\eta_1$, the D$(-1,2)$-brane has $\epsilon_1=+1$,
$\epsilon_2=\eta_1$, $\theta_5=\pi(\eta_2-\eta_1)/2$, $\theta_6=0$ and 
the \db$(-1,2)$-brane has $\epsilon_1=-1$,
$\epsilon_2=\eta_1$, $\theta_5=\pi(\eta_2-\eta_1)/2$, $\theta_6=\pi$ 
(\cf equations~(\ref{truncbrane}),~(\ref{fractboun})).}}
\label{fig1}
\end{figure}

Consider first a \dh$(-1,1)$-brane stretching along $x^5$, say. This is 
stable for all values of $R_5$ and for $R_i\ge 1/\sqrt{2}$ with 
$i=6,7,8$. For $R_i\le 1/\sqrt{2}$ the brane decays into a fractional 
$(-1,2)$ brane-anti-brane pair stretching along $x^5$ and $x^i$ in the 
same way as in the orbifold as shown in Figure~\ref{fig1}.\footnote{The fractional branes have opposite 
bulk \RR charge (as a result the object is non-BPS) as well as two of the 
four twisted \RR charges (this corresponds to turning on a Wilson line on 
one of the two branes). The other two twisted \RR charges are the same on 
each of the fractional branes.} From Table~\ref{tb1} the normalisations 
of the truncated brane's boundary states are
\be
{\cal N}^2_{U,(-1,1)}=\frac{1}{64}\frac{R_5}{R_iR_jR_k}
\,,\qquad {\cal N}^2_{T,(-1,1)}=\frac{8}{64}\,,
\ee
while each of the fractional branes is normalised as
\be
{\cal N}^2_{U,(-1,2)}=\frac{1}{128}\frac{R_5R_i}{R_jR_k}
\,,\qquad {\cal N}^2_{T,(-1,2)}=\frac{4}{128}\,.
\ee
Since the untwisted (twisted) sector normalisation ${\cal N}_U$ (${\cal 
N}_T$) is proportional to the mass (twisted \RR charge) of a brane one 
can easily verify that at the critical radius $R_i=1/\sqrt{2}$ the 
truncated brane has the same mass and charges as the pair of fractional 
branes. 

\begin{figure}[htb]
\begin{center}
\begin{picture}(0,0)%
\includegraphics{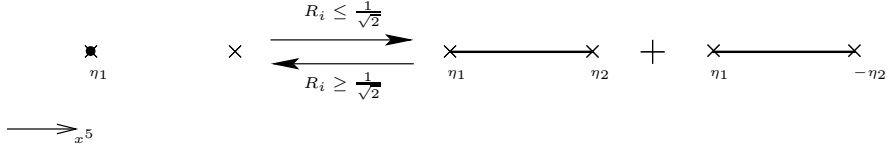}%
\end{picture}%
\setlength{\unitlength}{3947sp}%
\begingroup\makeatletter\ifx\SetFigFont\undefined%
\gdef\SetFigFont#1#2#3#4#5{%
  \reset@font\fontsize{#1}{#2pt}%
  \fontfamily{#3}\fontseries{#4}\fontshape{#5}%
  \selectfont}%
\fi\endgroup%
\begin{picture}(5384,872)(284,-111)
\put(826,314){\makebox(0,0)[lb]{\smash{\SetFigFont{5}{6.0}{\rmdefault}{\mddefault}{\updefault}{\color[rgb]{0,0,0}$\eta_1$}%
}}}
\put(731,-111){\makebox(0,0)[lb]{\smash{\SetFigFont{5}{6.0}{\rmdefault}{\mddefault}{\updefault}{\color[rgb]{0,0,0}$x^5$}%
}}}
\put(5626,314){\makebox(0,0)[lb]{\smash{\SetFigFont{5}{6.0}{\rmdefault}{\mddefault}{\updefault}{\color[rgb]{0,0,0}$-\eta_2$}%
}}}
\put(4726,314){\makebox(0,0)[lb]{\smash{\SetFigFont{5}{6.0}{\rmdefault}{\mddefault}{\updefault}{\color[rgb]{0,0,0}$\eta_1$}%
}}}
\put(3976,314){\makebox(0,0)[lb]{\smash{\SetFigFont{5}{6.0}{\rmdefault}{\mddefault}{\updefault}{\color[rgb]{0,0,0}$\eta_2$}%
}}}
\put(3076,314){\makebox(0,0)[lb]{\smash{\SetFigFont{5}{6.0}{\rmdefault}{\mddefault}{\updefault}{\color[rgb]{0,0,0}$\eta_1$}%
}}}
\put(2176,239){\makebox(0,0)[lb]{\smash{\SetFigFont{6}{7.2}{\rmdefault}{\mddefault}{\updefault}{\color[rgb]{0,0,0}$R_i\ge\frac{1}{\sqrt{2}}$}%
}}}
\put(2176,689){\makebox(0,0)[lb]{\smash{\SetFigFont{6}{7.2}{\rmdefault}{\mddefault}{\updefault}{\color[rgb]{0,0,0}$R_i\le\frac{1}{\sqrt{2}}$}%
}}}
\end{picture}
\end{center}
\caption{{\scriptsize The decay of a 
truncated (-1,0)-brane into a pair of \dh(-1,1)-branes. $\eta_i=\pm 1$ are the
signs of the twisted \RR charges at the fixed points indicated by the crosses.
The \dh-branes have $\epsilon=\eta$, with opposite Wilson lines $\theta_5$ on the
\dh$(-1,1)$-branes.}}
\label{fig2}
\end{figure}

Similarly, a \dh$(-1,0)$-brane is unstable for $R_i\le 1/\sqrt{2}$; it 
decays into a pair of \dh$(-1,1)$-branes which stretch along $x^i$ as shown in Figure~\ref{fig2}. The 
two \dh$(-1,1)$-branes have the same twisted \RR charge at the fixed 
point at which the original \dh$(-1,0)$-brane was located and opposite 
twisted \RR charges at the other fixed point. Note that unlike a single 
\dh$(-1,1)$-brane, due to non-trivial Chan-Paton factors a pair of 
\dh$(-1,1)$-branes does indeed develop a tachyon from open-string 
momentum modes. Due to the relative Wilson line open strings with an endpoint on 
each of the two \dh$(-1,1)$-branes have half-integral momentum states 
which become tachyonic for
\be
R_i\ge\frac{1}{\sqrt{2}}
\ee
and the configuration decays back into a \dh$(-1,0)$-brane.

Turning to the $r=3$ truncated branes we note that a \dh$(3,1)$-brane 
decays into a pair of fractional $(3,2)$-branes of opposite bulk charges 
(with one of the fractional branes having a Wilson line so as to conserve 
twisted \RR charge in the decay) for $R_\perp\le 1/\sqrt{2}$. However, 
the $(3,2)$-brane has $2\times 2$ Chan-Paton factors since it is a 
5-brane~\cite{GP}. This makes it twice as massive and carry twice the 
twisted \RR charge.\footnote{This guarantees that the fractional $(-1,2)$ 
and $(3,2)$-branes satisfy the Dirac quantisation condition.} In order to 
conserve energy and charge in this decay the $(3,1)$-brane has also got 
to have $2\times 2$ Chan-Paton factors and as such its stability is 
given in equation~(\ref{orbstab}).
We hope to present a more detailed analysis of this and some of the other 
decay channels of the $r=3$ truncated branes in the future.

\section{Torsion charged D-branes}\label{sec4}
\setcounter{equation}{0}

Torsion charged branes do not couple to \RR sectors. They have previously 
been encountered in Type I-like theories~\cite{Snbps,WittK,Frau,DS}, shown to carry $\Z_2$ charges 
(since a pair of them decays into the vacuum) and couple only to the 
untwisted \NSNS sector. In the first part of this section we find a new 
class of torsion branes coupling to both twisted and untwisted \NSNS 
sectors. In the decompactified theory these are $\Z_2\oplus\Z_2$ charged. 
In the second part of this section we find $\Z_2$-charged branes in the GP model 
which couple only to the untwisted \NSNS sector much like the torsion 
branes of Type I. All the torsion branes in the GP orientifold give rise 
to open string tachyons between them and the tadpole canceling $(5,0)$- 
or $(5,4)$-branes. Nonetheless this shows that, while the branes are 
unstable, the charges they carry are present in the theory. As we will see 
in section~\ref{sec7} in the DPBZ 
model, similar torsion branes do not have such tachyons and so are 
genuinely stable.

\subsection{Torsion branes with twisted \NSNS couplings}

Consider an $(r,2)$-brane of the form
\be
\ket{D(r,2)} =  {\cal N}_{(r,2),U}
\left|B(r,2)\right>_{\mbox{\scriptsize\NSNS}} 
+ \epsilon \; {\cal N}_{(r,2),T} \sum_{\alpha=1}^{4}
e^{i\theta_\alpha}
\left|B(r,2)\right>_{\mbox{\scriptsize\NSNS,T$_\alpha$}}\,,
\label{torbrane}
\ee
where $\epsilon=\pm 1$, $\alpha$ labels the different fixed points 
between which the
brane stretches, and $\theta_\alpha=0,\pi$ is the Wilson line that is 
associated to the difference of the fixed point $\alpha$ and the origin.\footnote{In 
particular $\theta_1=0$, 
$\theta_2=\theta_5$, $\theta_3=\theta_6$ and $\theta_4=\theta_5+\theta_6$ where 
$\theta_5,\theta_6$ are the 
Wilson lines on the brane in the $x^5,x^6$ directions.} 
We keep $r$ and the normalisations ${\cal N}$ unspecified.\footnote{We 
take $s=2$ as this is the only value of $s$ for which the twisted \NSNS 
boundary state is $\Omega$, GSO, and $\I_4$ invariant (see 
Appendix~\ref{appb} for details).} The one-loop partition function for an 
open string ending on such a brane is computed in Appendix~\ref{appd}; 
one finds that for $r=4,5$ and suitable normalisations
the ground-state tachyon is projected out stabilising the $(4,2)$ and 
$(5,2)$-branes. The $(4,2)$-brane is a D6-brane of Type I fixed under $\I_4$. Such
D-branes have been shown to be inconsistent in Type I~\cite{WittK,Frau} and so are
inconsistent in the present orientifold. Expanding further in winding and 
momenta one finds that for
\be
\frac{1}{R_{||}^2}+R_\perp^2\ge\half
\label{condtortor}
\ee
the $(5,2)$-branes are stable. In particular in the non-compact orbifold the 
branes are always stable. Similar stability conditions have been 
previously encountered for integrally charged branes in~\cite{SCY,MajSen}. 

Since D-branes are described by K-theory, a complete set of branes 
transverse to a particular sub-manifold of spacetime has to form a group. 
In the non-compact theory we have found two branes with
$\epsilon=\pm 1$ to which we will refer to as $g_\pm$.\footnote{In the 
non-compact theory there is only one \NSNS twisted sector and the D-brane 
boundary state is 
\ba
\ket{D_{\Z_2\oplus\Z_2}(r,2)} =  {\cal N}_{(r,2),U}
\left|B(r,2)\right>_{\mbox{\scriptsize\NSNS}} 
+ \epsilon \; {\cal N}_{(r,2),T} 
\left|B(r,2)\right>_{\mbox{\scriptsize\NSNS,T}}\,.\nonumber
\ea} In the next subsection we show that an $\epsilon=+1$ and an 
$\epsilon=-1$ brane join to form another stable brane, call it $h$, 
transverse to the same sub-manifold
\be
g_+ + g_-=h\,,
\label{z2arg}
\ee
and further that this brane is $\Z_2$ charged
\be
h+h=1\,,
\label{ord2h}
\ee
where $1$ is the vacuum. K-theory tells us that $\{1,g_+,g_-,h\}$ form a 
group. Since each of the branes is different from the vacuum this can be 
$\Z_2\oplus\Z_2$ or $\Z_4$.\footnote{At first one might consider $\Z_2$ 
but equation~(\ref{z2arg}) would imply that one of $g_+,g_-$ would equal 
$1$, the vacuum.} $\Z_4$ is impossible: for if that were the group 
equation~(\ref{ord2h}) implies that $h=x^2$, where $x$ is the generator 
of $\Z_4$, and~(\ref{z2arg}) gives $g_+=g_-=x$ (or $g_+=g_-=x^3$). There 
should then be another brane corresponding to $x^3$ (or $x$), and since 
the complete list of branes is $\{1,g_+,g_-,h\}$ this is impossible. 
Hence the torsion branes in the decompactified theory are charged under 
$\Z_2\oplus\Z_2$. As a result we have further learnt that the unstable 
configurations $g_++g_+$ and $g_-+g_-$ both decay into the vacuum.

\begin{figure}[htb]
\begin{center}
\begin{picture}(0,0)%
\includegraphics{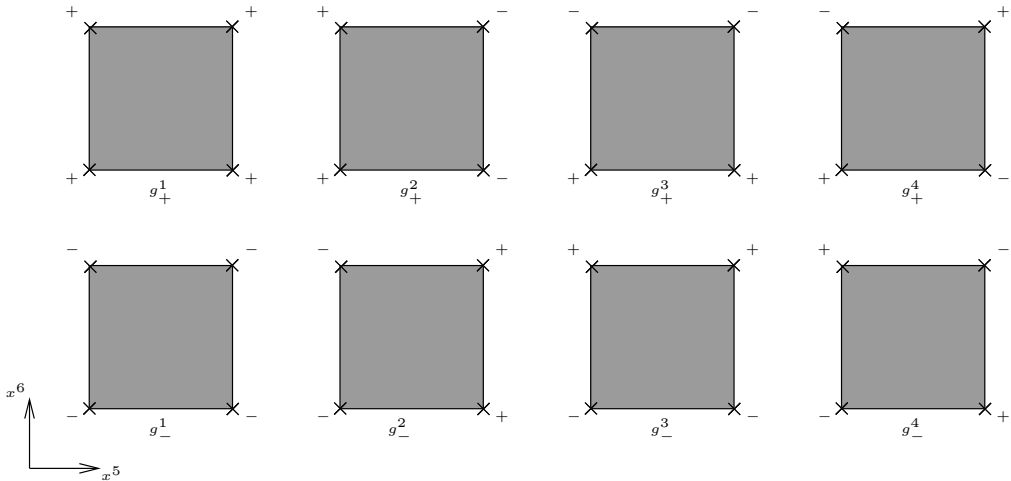}%
\end{picture}%
\setlength{\unitlength}{3947sp}%
\begingroup\makeatletter\ifx\SetFigFont\undefined%
\gdef\SetFigFont#1#2#3#4#5{%
  \reset@font\fontsize{#1}{#2pt}%
  \fontfamily{#3}\fontseries{#4}\fontshape{#5}%
  \selectfont}%
\fi\endgroup%
\begin{picture}(6225,3000)(1,-2311)
\put(3526,-436){\makebox(0,0)[lb]{\smash{\SetFigFont{5}{6.0}{\rmdefault}{\mddefault}{\updefault}{\color[rgb]{0,0,0}$+$}%
}}}
\put(4651,-436){\makebox(0,0)[lb]{\smash{\SetFigFont{5}{6.0}{\rmdefault}{\mddefault}{\updefault}{\color[rgb]{0,0,0}$+$}%
}}}
\put(3076,-436){\makebox(0,0)[lb]{\smash{\SetFigFont{5}{6.0}{\rmdefault}{\mddefault}{\updefault}{\color[rgb]{0,0,0}$-$}%
}}}
\put(1951,614){\makebox(0,0)[lb]{\smash{\SetFigFont{5}{6.0}{\rmdefault}{\mddefault}{\updefault}{\color[rgb]{0,0,0}$+$}%
}}}
\put(1951,-436){\makebox(0,0)[lb]{\smash{\SetFigFont{5}{6.0}{\rmdefault}{\mddefault}{\updefault}{\color[rgb]{0,0,0}$+$}%
}}}
\put(376,614){\makebox(0,0)[lb]{\smash{\SetFigFont{5}{6.0}{\rmdefault}{\mddefault}{\updefault}{\color[rgb]{0,0,0}$+$}%
}}}
\put(376,-436){\makebox(0,0)[lb]{\smash{\SetFigFont{5}{6.0}{\rmdefault}{\mddefault}{\updefault}{\color[rgb]{0,0,0}$+$}%
}}}
\put(1501,-436){\makebox(0,0)[lb]{\smash{\SetFigFont{5}{6.0}{\rmdefault}{\mddefault}{\updefault}{\color[rgb]{0,0,0}$+$}%
}}}
\put(5101,-436){\makebox(0,0)[lb]{\smash{\SetFigFont{5}{6.0}{\rmdefault}{\mddefault}{\updefault}{\color[rgb]{0,0,0}$+$}%
}}}
\put(5101,614){\makebox(0,0)[lb]{\smash{\SetFigFont{5}{6.0}{\rmdefault}{\mddefault}{\updefault}{\color[rgb]{0,0,0}$-$}%
}}}
\put(6226,-436){\makebox(0,0)[lb]{\smash{\SetFigFont{5}{6.0}{\rmdefault}{\mddefault}{\updefault}{\color[rgb]{0,0,0}$-$}%
}}}
\put(3526,614){\makebox(0,0)[lb]{\smash{\SetFigFont{5}{6.0}{\rmdefault}{\mddefault}{\updefault}{\color[rgb]{0,0,0}$-$}%
}}}
\put(601,-2311){\makebox(0,0)[lb]{\smash{\SetFigFont{5}{6.0}{\rmdefault}{\mddefault}{\updefault}{\color[rgb]{0,0,0}$x^5$}%
}}}
\put(5101,-886){\makebox(0,0)[lb]{\smash{\SetFigFont{5}{6.0}{\rmdefault}{\mddefault}{\updefault}{\color[rgb]{0,0,0}$+$}%
}}}
\put(376,-886){\makebox(0,0)[lb]{\smash{\SetFigFont{5}{6.0}{\rmdefault}{\mddefault}{\updefault}{\color[rgb]{0,0,0}$-$}%
}}}
\put(1501,-1936){\makebox(0,0)[lb]{\smash{\SetFigFont{5}{6.0}{\rmdefault}{\mddefault}{\updefault}{\color[rgb]{0,0,0}$-$}%
}}}
\put(376,-1936){\makebox(0,0)[lb]{\smash{\SetFigFont{5}{6.0}{\rmdefault}{\mddefault}{\updefault}{\color[rgb]{0,0,0}$-$}%
}}}
\put(  1,-1786){\makebox(0,0)[lb]{\smash{\SetFigFont{5}{6.0}{\rmdefault}{\mddefault}{\updefault}{\color[rgb]{0,0,0}$x^6$}%
}}}
\put(1951,-1936){\makebox(0,0)[lb]{\smash{\SetFigFont{5}{6.0}{\rmdefault}{\mddefault}{\updefault}{\color[rgb]{0,0,0}$-$}%
}}}
\put(1951,-886){\makebox(0,0)[lb]{\smash{\SetFigFont{5}{6.0}{\rmdefault}{\mddefault}{\updefault}{\color[rgb]{0,0,0}$-$}%
}}}
\put(3076,-1936){\makebox(0,0)[lb]{\smash{\SetFigFont{5}{6.0}{\rmdefault}{\mddefault}{\updefault}{\color[rgb]{0,0,0}$+$}%
}}}
\put(3526,-1936){\makebox(0,0)[lb]{\smash{\SetFigFont{5}{6.0}{\rmdefault}{\mddefault}{\updefault}{\color[rgb]{0,0,0}$-$}%
}}}
\put(4651,-1936){\makebox(0,0)[lb]{\smash{\SetFigFont{5}{6.0}{\rmdefault}{\mddefault}{\updefault}{\color[rgb]{0,0,0}$-$}%
}}}
\put(3526,-886){\makebox(0,0)[lb]{\smash{\SetFigFont{5}{6.0}{\rmdefault}{\mddefault}{\updefault}{\color[rgb]{0,0,0}$+$}%
}}}
\put(5101,-1936){\makebox(0,0)[lb]{\smash{\SetFigFont{5}{6.0}{\rmdefault}{\mddefault}{\updefault}{\color[rgb]{0,0,0}$-$}%
}}}
\put(6226,-1936){\makebox(0,0)[lb]{\smash{\SetFigFont{5}{6.0}{\rmdefault}{\mddefault}{\updefault}{\color[rgb]{0,0,0}$+$}%
}}}
\put(6226,614){\makebox(0,0)[lb]{\smash{\SetFigFont{5}{6.0}{\rmdefault}{\mddefault}{\updefault}{\color[rgb]{0,0,0}$+$}%
}}}
\put(901,-511){\makebox(0,0)[lb]{\smash{\SetFigFont{5}{6.0}{\rmdefault}{\mddefault}{\updefault}{\color[rgb]{0,0,0}$g_+^1$}%
}}}
\put(1501,614){\makebox(0,0)[lb]{\smash{\SetFigFont{5}{6.0}{\rmdefault}{\mddefault}{\updefault}{\color[rgb]{0,0,0}$+$}%
}}}
\put(2476,-511){\makebox(0,0)[lb]{\smash{\SetFigFont{5}{6.0}{\rmdefault}{\mddefault}{\updefault}{\color[rgb]{0,0,0}$g_+^2$}%
}}}
\put(1501,-886){\makebox(0,0)[lb]{\smash{\SetFigFont{5}{6.0}{\rmdefault}{\mddefault}{\updefault}{\color[rgb]{0,0,0}$-$}%
}}}
\put(901,-2011){\makebox(0,0)[lb]{\smash{\SetFigFont{5}{6.0}{\rmdefault}{\mddefault}{\updefault}{\color[rgb]{0,0,0}$g_-^1$}%
}}}
\put(3076,614){\makebox(0,0)[lb]{\smash{\SetFigFont{5}{6.0}{\rmdefault}{\mddefault}{\updefault}{\color[rgb]{0,0,0}$-$}%
}}}
\put(4051,-511){\makebox(0,0)[lb]{\smash{\SetFigFont{5}{6.0}{\rmdefault}{\mddefault}{\updefault}{\color[rgb]{0,0,0}$g_+^3$}%
}}}
\put(3076,-886){\makebox(0,0)[lb]{\smash{\SetFigFont{5}{6.0}{\rmdefault}{\mddefault}{\updefault}{\color[rgb]{0,0,0}$+$}%
}}}
\put(2401,-2011){\makebox(0,0)[lb]{\smash{\SetFigFont{5}{6.0}{\rmdefault}{\mddefault}{\updefault}{\color[rgb]{0,0,0}$g_-^2$}%
}}}
\put(4651,-886){\makebox(0,0)[lb]{\smash{\SetFigFont{5}{6.0}{\rmdefault}{\mddefault}{\updefault}{\color[rgb]{0,0,0}$+$}%
}}}
\put(4651,614){\makebox(0,0)[lb]{\smash{\SetFigFont{5}{6.0}{\rmdefault}{\mddefault}{\updefault}{\color[rgb]{0,0,0}$-$}%
}}}
\put(4051,-2011){\makebox(0,0)[lb]{\smash{\SetFigFont{5}{6.0}{\rmdefault}{\mddefault}{\updefault}{\color[rgb]{0,0,0}$g_-^3$}%
}}}
\put(5626,-2011){\makebox(0,0)[lb]{\smash{\SetFigFont{5}{6.0}{\rmdefault}{\mddefault}{\updefault}{\color[rgb]{0,0,0}$g_-^4$}%
}}}
\put(5626,-511){\makebox(0,0)[lb]{\smash{\SetFigFont{5}{6.0}{\rmdefault}{\mddefault}{\updefault}{\color[rgb]{0,0,0}$g_+^4$}%
}}}
\put(6226,-886){\makebox(0,0)[lb]{\smash{\SetFigFont{5}{6.0}{\rmdefault}{\mddefault}{\updefault}{\color[rgb]{0,0,0}$-$}%
}}}
\end{picture}
\end{center}
\caption{{\scriptsize The eight $(r,2)$-branes 
($g^1_\pm,\dots ,g^4_\pm$) coupling to both the twisted and untwisted \NSNS sectors 
with no constant gauge field in the internal directions
The configurations correspond to the choices $\epsilon=\pm 1$ and $\theta_5,\theta_6=0,\pi$ 
in the boundary state~(\ref{torbrane}). In the compactified space the branes are taken to 
stretch along $x^5$ and $x^6$ and the figure shows only these directions.}}
\label{fig3}
\end{figure}

\begin{figure}[htb]
\begin{center}
\begin{picture}(0,0)%
\includegraphics{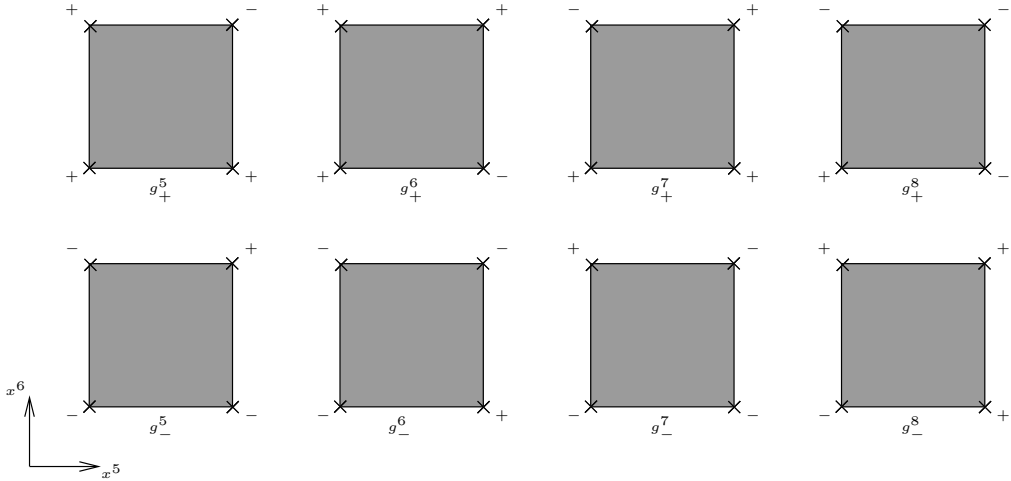}%
\end{picture}%
\setlength{\unitlength}{3947sp}%
\begingroup\makeatletter\ifx\SetFigFont\undefined%
\gdef\SetFigFont#1#2#3#4#5{%
  \reset@font\fontsize{#1}{#2pt}%
  \fontfamily{#3}\fontseries{#4}\fontshape{#5}%
  \selectfont}%
\fi\endgroup%
\begin{picture}(6225,3000)(1,-2311)
\put(3526,-886){\makebox(0,0)[lb]{\smash{\SetFigFont{5}{6.0}{\rmdefault}{\mddefault}{\updefault}{\color[rgb]{0,0,0}$+$}%
}}}
\put(4651,-1936){\makebox(0,0)[lb]{\smash{\SetFigFont{5}{6.0}{\rmdefault}{\mddefault}{\updefault}{\color[rgb]{0,0,0}$-$}%
}}}
\put(3526,-1936){\makebox(0,0)[lb]{\smash{\SetFigFont{5}{6.0}{\rmdefault}{\mddefault}{\updefault}{\color[rgb]{0,0,0}$-$}%
}}}
\put(3076,-1936){\makebox(0,0)[lb]{\smash{\SetFigFont{5}{6.0}{\rmdefault}{\mddefault}{\updefault}{\color[rgb]{0,0,0}$+$}%
}}}
\put(601,-2311){\makebox(0,0)[lb]{\smash{\SetFigFont{5}{6.0}{\rmdefault}{\mddefault}{\updefault}{\color[rgb]{0,0,0}$x^5$}%
}}}
\put(376,-886){\makebox(0,0)[lb]{\smash{\SetFigFont{5}{6.0}{\rmdefault}{\mddefault}{\updefault}{\color[rgb]{0,0,0}$-$}%
}}}
\put(1501,-1936){\makebox(0,0)[lb]{\smash{\SetFigFont{5}{6.0}{\rmdefault}{\mddefault}{\updefault}{\color[rgb]{0,0,0}$-$}%
}}}
\put(376,-1936){\makebox(0,0)[lb]{\smash{\SetFigFont{5}{6.0}{\rmdefault}{\mddefault}{\updefault}{\color[rgb]{0,0,0}$-$}%
}}}
\put(  1,-1786){\makebox(0,0)[lb]{\smash{\SetFigFont{5}{6.0}{\rmdefault}{\mddefault}{\updefault}{\color[rgb]{0,0,0}$x^6$}%
}}}
\put(1951,-886){\makebox(0,0)[lb]{\smash{\SetFigFont{5}{6.0}{\rmdefault}{\mddefault}{\updefault}{\color[rgb]{0,0,0}$-$}%
}}}
\put(1951,-1936){\makebox(0,0)[lb]{\smash{\SetFigFont{5}{6.0}{\rmdefault}{\mddefault}{\updefault}{\color[rgb]{0,0,0}$-$}%
}}}
\put(5101,-886){\makebox(0,0)[lb]{\smash{\SetFigFont{5}{6.0}{\rmdefault}{\mddefault}{\updefault}{\color[rgb]{0,0,0}$+$}%
}}}
\put(6226,-436){\makebox(0,0)[lb]{\smash{\SetFigFont{5}{6.0}{\rmdefault}{\mddefault}{\updefault}{\color[rgb]{0,0,0}$-$}%
}}}
\put(5101,614){\makebox(0,0)[lb]{\smash{\SetFigFont{5}{6.0}{\rmdefault}{\mddefault}{\updefault}{\color[rgb]{0,0,0}$-$}%
}}}
\put(6226,-1936){\makebox(0,0)[lb]{\smash{\SetFigFont{5}{6.0}{\rmdefault}{\mddefault}{\updefault}{\color[rgb]{0,0,0}$+$}%
}}}
\put(5101,-1936){\makebox(0,0)[lb]{\smash{\SetFigFont{5}{6.0}{\rmdefault}{\mddefault}{\updefault}{\color[rgb]{0,0,0}$-$}%
}}}
\put(1501,-436){\makebox(0,0)[lb]{\smash{\SetFigFont{5}{6.0}{\rmdefault}{\mddefault}{\updefault}{\color[rgb]{0,0,0}$+$}%
}}}
\put(376,-436){\makebox(0,0)[lb]{\smash{\SetFigFont{5}{6.0}{\rmdefault}{\mddefault}{\updefault}{\color[rgb]{0,0,0}$+$}%
}}}
\put(376,614){\makebox(0,0)[lb]{\smash{\SetFigFont{5}{6.0}{\rmdefault}{\mddefault}{\updefault}{\color[rgb]{0,0,0}$+$}%
}}}
\put(1951,-436){\makebox(0,0)[lb]{\smash{\SetFigFont{5}{6.0}{\rmdefault}{\mddefault}{\updefault}{\color[rgb]{0,0,0}$+$}%
}}}
\put(1951,614){\makebox(0,0)[lb]{\smash{\SetFigFont{5}{6.0}{\rmdefault}{\mddefault}{\updefault}{\color[rgb]{0,0,0}$+$}%
}}}
\put(3076,-436){\makebox(0,0)[lb]{\smash{\SetFigFont{5}{6.0}{\rmdefault}{\mddefault}{\updefault}{\color[rgb]{0,0,0}$-$}%
}}}
\put(4651,-436){\makebox(0,0)[lb]{\smash{\SetFigFont{5}{6.0}{\rmdefault}{\mddefault}{\updefault}{\color[rgb]{0,0,0}$+$}%
}}}
\put(3526,-436){\makebox(0,0)[lb]{\smash{\SetFigFont{5}{6.0}{\rmdefault}{\mddefault}{\updefault}{\color[rgb]{0,0,0}$+$}%
}}}
\put(3526,614){\makebox(0,0)[lb]{\smash{\SetFigFont{5}{6.0}{\rmdefault}{\mddefault}{\updefault}{\color[rgb]{0,0,0}$-$}%
}}}
\put(5101,-436){\makebox(0,0)[lb]{\smash{\SetFigFont{5}{6.0}{\rmdefault}{\mddefault}{\updefault}{\color[rgb]{0,0,0}$+$}%
}}}
\put(901,-2011){\makebox(0,0)[lb]{\smash{\SetFigFont{5}{6.0}{\rmdefault}{\mddefault}{\updefault}{\color[rgb]{0,0,0}$g_-^5$}%
}}}
\put(2401,-2011){\makebox(0,0)[lb]{\smash{\SetFigFont{5}{6.0}{\rmdefault}{\mddefault}{\updefault}{\color[rgb]{0,0,0}$g_-^6$}%
}}}
\put(901,-511){\makebox(0,0)[lb]{\smash{\SetFigFont{5}{6.0}{\rmdefault}{\mddefault}{\updefault}{\color[rgb]{0,0,0}$g_+^5$}%
}}}
\put(3076,-886){\makebox(0,0)[lb]{\smash{\SetFigFont{5}{6.0}{\rmdefault}{\mddefault}{\updefault}{\color[rgb]{0,0,0}$-$}%
}}}
\put(1501,614){\makebox(0,0)[lb]{\smash{\SetFigFont{5}{6.0}{\rmdefault}{\mddefault}{\updefault}{\color[rgb]{0,0,0}$-$}%
}}}
\put(2476,-511){\makebox(0,0)[lb]{\smash{\SetFigFont{5}{6.0}{\rmdefault}{\mddefault}{\updefault}{\color[rgb]{0,0,0}$g_+^6$}%
}}}
\put(4051,-2011){\makebox(0,0)[lb]{\smash{\SetFigFont{5}{6.0}{\rmdefault}{\mddefault}{\updefault}{\color[rgb]{0,0,0}$g_-^7$}%
}}}
\put(4651,-886){\makebox(0,0)[lb]{\smash{\SetFigFont{5}{6.0}{\rmdefault}{\mddefault}{\updefault}{\color[rgb]{0,0,0}$-$}%
}}}
\put(3076,614){\makebox(0,0)[lb]{\smash{\SetFigFont{5}{6.0}{\rmdefault}{\mddefault}{\updefault}{\color[rgb]{0,0,0}$+$}%
}}}
\put(4051,-511){\makebox(0,0)[lb]{\smash{\SetFigFont{5}{6.0}{\rmdefault}{\mddefault}{\updefault}{\color[rgb]{0,0,0}$g_+^7$}%
}}}
\put(5626,-2011){\makebox(0,0)[lb]{\smash{\SetFigFont{5}{6.0}{\rmdefault}{\mddefault}{\updefault}{\color[rgb]{0,0,0}$g_-^8$}%
}}}
\put(6226,614){\makebox(0,0)[lb]{\smash{\SetFigFont{5}{6.0}{\rmdefault}{\mddefault}{\updefault}{\color[rgb]{0,0,0}$-$}%
}}}
\put(4651,614){\makebox(0,0)[lb]{\smash{\SetFigFont{5}{6.0}{\rmdefault}{\mddefault}{\updefault}{\color[rgb]{0,0,0}$+$}%
}}}
\put(5626,-511){\makebox(0,0)[lb]{\smash{\SetFigFont{5}{6.0}{\rmdefault}{\mddefault}{\updefault}{\color[rgb]{0,0,0}$g_+^8$}%
}}}
\put(6226,-886){\makebox(0,0)[lb]{\smash{\SetFigFont{5}{6.0}{\rmdefault}{\mddefault}{\updefault}{\color[rgb]{0,0,0}$+$}%
}}}
\put(1501,-886){\makebox(0,0)[lb]{\smash{\SetFigFont{5}{6.0}{\rmdefault}{\mddefault}{\updefault}{\color[rgb]{0,0,0}$+$}%
}}}
\end{picture}
\end{center}
\caption{{\scriptsize The eight $(5,2)$-branes ($g^5_\pm,\dots ,g^8_\pm$) 
coupling to both the twisted and untwisted \NSNS sectors with a 
constant magnetic flux $F_{56}=\pm 1$. 
The configurations correspond to the choices $\epsilon=\pm 1$ and $\theta_5,\theta_6=0,\pi$ 
in the boundary state~(\ref{torbrane}). Each is stable for a particular $F_{56}$. 
In the compactified space the branes are taken to 
stretch along $x^5$ and $x^6$ and the figure shows only these directions.}}
\label{fig4}
\end{figure}

In the compact theory there are eight branes with twisted \NSNS couplings 
transverse to the same spacetime sub-manifold corresponding to the choices 
associated with $\epsilon$ and the two Wilson lines $\theta_i$. Further by turning on a constant 
gauge field $F_{ij}=\pm 1$ in the two internal directions along which the brane stretches,
we obtain eight more stable configurations. This is quite analogous to~\cite{ABG} (see also~\cite{RU} for
related orientifolds) where 
it was used to describe a certain decay product of a tadpole canceling brane-anti-brane pair into a
truncated brane with a constant magnetic flux.\footnote{We are grateful to M.R. Gaberdiel 
for bringing this construction to our attention and for suggesting its application here.}
We refer to these sixteen branes as $g^i_\pm$ as shown in Figures~\ref{fig4} and~\ref{fig5}. In the 
next subsection we show that as in the decompactified case there is a 
brane $h$ such that
\be
g_+^i+g_-^i=h\,,
\label{z2comparg}
\ee
for all $i$ and $h$ is again order two~(\ref{ord2h}). It would be interesting to determine the
overall group structure in the compact case.

Before turning to the second class of torsion branes we mention a 
particularly useful way to think about the $(5,2)$-brane. This brane is 
an $\Omega$-invariant configuration of two fractional $(5,2)$-branes in 
the $\I_4$ orbifold with opposite bulk and twisted \RR charges. Such a 
configuration condenses into the vacuum in the orbifold; in the 
orientifold however there is a torsion charge 
that this pair carries which stabilises it. 

\subsection{Torsion branes with no twisted \NSNS couplings}

In this section we consider a D-brane coupling only to the untwisted 
\NSNS sector
\be
\ket{D_{\ms{$\Z_2$}}(r,s)}=\ket{B(r,s)}_{\ms{\NSNS}}\,,
\label{z2br}
\ee
with arbitrary $r$, $s$, and normalisation. From the one-loop partition 
function computed in Appendix~\ref{appd} one finds that for $r=4,5$ and 
$s=1,2,3$ the ground-state tachyon is removed. We will show that the $(r,1)$-,
$(r,2)$- and $(r,3)$-branes decay into one another. Since the $(4,2)$-brane is inconsistent
(see previous subsection) we expect the $(4,1)$- and $(4,3)$-branes to be also inconsitent.  
The normalisations (listed in 
Appendix~\ref{appd}) for the $s=2$ branes indeed make them bound states 
of two branes from the previous subsection as described in 
equations~(\ref{z2arg}) and~(\ref{z2comparg}).

The $(5,s)$-branes with no twisted \NSNS couplings are stable for
\be
R_i\le\sqrt{2}\,,i=5,\dots,4+s\,,\qquad 
R_j\ge\frac{1}{\sqrt{2}}\,,j=5+s,\dots,8\,,
\label{full}
\ee
where we have taken the brane to lie along $x^5,\dots,x^{4+s}$ and be 
transverse to $x^{5+s},\dots,x^8$. 
\begin{figure}[htb]
\begin{center}
\begin{picture}(0,0)%
\includegraphics{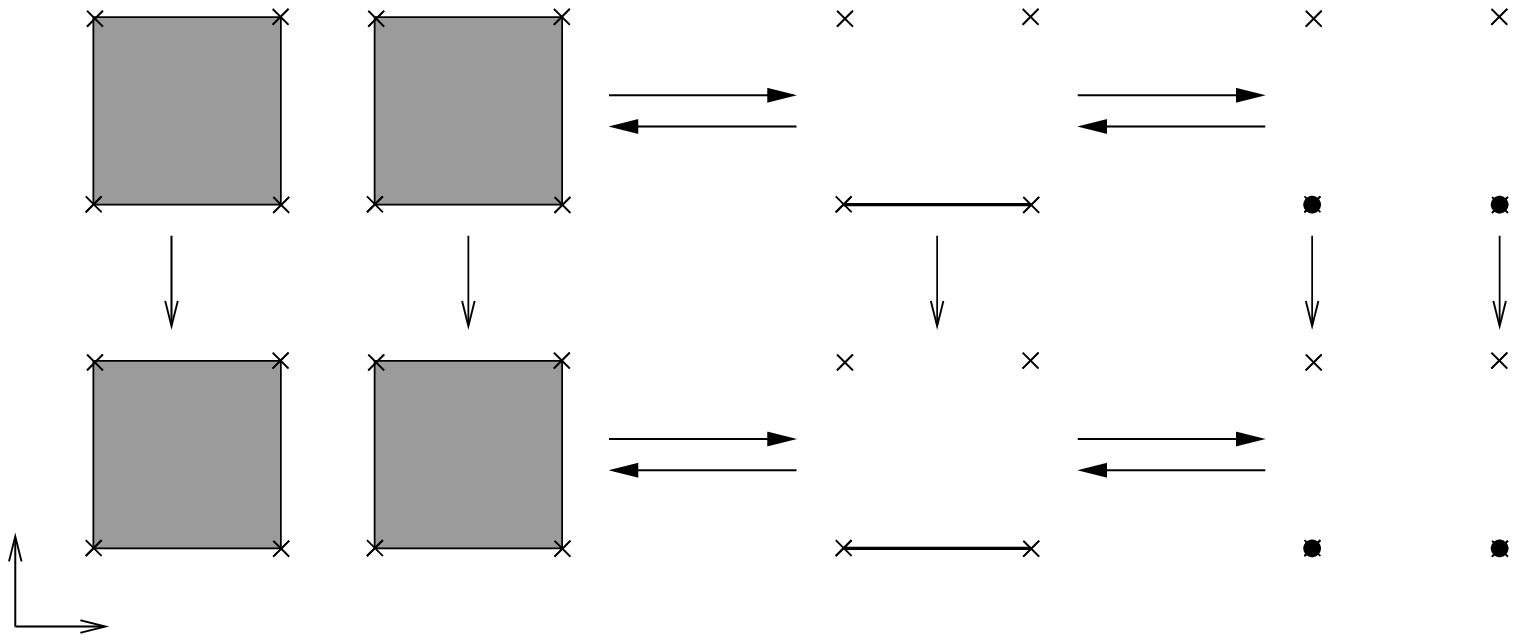}%
\end{picture}%
\setlength{\unitlength}{3947sp}%
\begingroup\makeatletter\ifx\SetFigFont\undefined%
\gdef\SetFigFont#1#2#3#4#5{%
  \reset@font\fontsize{#1}{#2pt}%
  \fontfamily{#3}\fontseries{#4}\fontshape{#5}%
  \selectfont}%
\fi\endgroup%
\begin{picture}(7350,3525)(1,-2836)
\put(1726,-1036){\makebox(0,0)[lb]{\smash{\SetFigFont{5}{6.0}{\rmdefault}{\mddefault}{\updefault}{\color[rgb]{0,0,0}$+$}%
}}}
\put(3976,-2086){\makebox(0,0)[lb]{\smash{\SetFigFont{5}{6.0}{\rmdefault}{\mddefault}{\updefault}{\color[rgb]{0,0,0}$-$}%
}}}
\put(2851,-2086){\makebox(0,0)[lb]{\smash{\SetFigFont{5}{6.0}{\rmdefault}{\mddefault}{\updefault}{\color[rgb]{0,0,0}$-$}%
}}}
\put(1501,-436){\makebox(0,0)[lb]{\smash{\SetFigFont{5}{6.0}{\rmdefault}{\mddefault}{\updefault}{\color[rgb]{0,0,0}$+$}%
}}}
\put(376,-436){\makebox(0,0)[lb]{\smash{\SetFigFont{5}{6.0}{\rmdefault}{\mddefault}{\updefault}{\color[rgb]{0,0,0}$+$}%
}}}
\put(376,614){\makebox(0,0)[lb]{\smash{\SetFigFont{5}{6.0}{\rmdefault}{\mddefault}{\updefault}{\color[rgb]{0,0,0}$+$}%
}}}
\put(601,-2461){\makebox(0,0)[lb]{\smash{\SetFigFont{5}{6.0}{\rmdefault}{\mddefault}{\updefault}{\color[rgb]{0,0,0}$x^5$}%
}}}
\put(376,-1036){\makebox(0,0)[lb]{\smash{\SetFigFont{5}{6.0}{\rmdefault}{\mddefault}{\updefault}{\color[rgb]{0,0,0}$-$}%
}}}
\put(1501,-2086){\makebox(0,0)[lb]{\smash{\SetFigFont{5}{6.0}{\rmdefault}{\mddefault}{\updefault}{\color[rgb]{0,0,0}$-$}%
}}}
\put(376,-2086){\makebox(0,0)[lb]{\smash{\SetFigFont{5}{6.0}{\rmdefault}{\mddefault}{\updefault}{\color[rgb]{0,0,0}$-$}%
}}}
\put(3976,-436){\makebox(0,0)[lb]{\smash{\SetFigFont{5}{6.0}{\rmdefault}{\mddefault}{\updefault}{\color[rgb]{0,0,0}$+$}%
}}}
\put(  1,-1936){\makebox(0,0)[lb]{\smash{\SetFigFont{5}{6.0}{\rmdefault}{\mddefault}{\updefault}{\color[rgb]{0,0,0}$x^6$}%
}}}
\put(7351,-436){\makebox(0,0)[lb]{\smash{\SetFigFont{5}{6.0}{\rmdefault}{\mddefault}{\updefault}{\color[rgb]{0,0,0}$+$}%
}}}
\put(7351,-2086){\makebox(0,0)[lb]{\smash{\SetFigFont{5}{6.0}{\rmdefault}{\mddefault}{\updefault}{\color[rgb]{0,0,0}$-$}%
}}}
\put(6226,-2086){\makebox(0,0)[lb]{\smash{\SetFigFont{5}{6.0}{\rmdefault}{\mddefault}{\updefault}{\color[rgb]{0,0,0}$-$}%
}}}
\put(6226,-436){\makebox(0,0)[lb]{\smash{\SetFigFont{5}{6.0}{\rmdefault}{\mddefault}{\updefault}{\color[rgb]{0,0,0}$+$}%
}}}
\put(1726,-436){\makebox(0,0)[lb]{\smash{\SetFigFont{5}{6.0}{\rmdefault}{\mddefault}{\updefault}{\color[rgb]{0,0,0}$+$}%
}}}
\put(2851,-436){\makebox(0,0)[lb]{\smash{\SetFigFont{5}{6.0}{\rmdefault}{\mddefault}{\updefault}{\color[rgb]{0,0,0}$-$}%
}}}
\put(1726,-2086){\makebox(0,0)[lb]{\smash{\SetFigFont{5}{6.0}{\rmdefault}{\mddefault}{\updefault}{\color[rgb]{0,0,0}$-$}%
}}}
\put(5101,-436){\makebox(0,0)[lb]{\smash{\SetFigFont{5}{6.0}{\rmdefault}{\mddefault}{\updefault}{\color[rgb]{0,0,0}$+$}%
}}}
\put(1726,614){\makebox(0,0)[lb]{\smash{\SetFigFont{5}{6.0}{\rmdefault}{\mddefault}{\updefault}{\color[rgb]{0,0,0}$-$}%
}}}
\put(5101,-2086){\makebox(0,0)[lb]{\smash{\SetFigFont{5}{6.0}{\rmdefault}{\mddefault}{\updefault}{\color[rgb]{0,0,0}$-$}%
}}}
\put(7351,-736){\makebox(0,0)[lb]{\smash{\SetFigFont{7}{8.4}{\rmdefault}{\mddefault}{\updefault}{\color[rgb]{0,0,0}$\Omega$}%
}}}
\put(7276,-1861){\makebox(0,0)[lb]{\smash{\SetFigFont{9}{10.8}{\rmdefault}{\mddefault}{\updefault}{\color[rgb]{0,0,0}$\bar{\mbox{D}}$}%
}}}
\put(6451,-736){\makebox(0,0)[lb]{\smash{\SetFigFont{7}{8.4}{\rmdefault}{\mddefault}{\updefault}{\color[rgb]{0,0,0}$\Omega$}%
}}}
\put(7276,-211){\makebox(0,0)[lb]{\smash{\SetFigFont{9}{10.8}{\rmdefault}{\mddefault}{\updefault}{\color[rgb]{0,0,0}$\bar{\mbox{D}}$}%
}}}
\put(4651,-736){\makebox(0,0)[lb]{\smash{\SetFigFont{7}{8.4}{\rmdefault}{\mddefault}{\updefault}{\color[rgb]{0,0,0}$\Omega$}%
}}}
\put(2851,-1036){\makebox(0,0)[lb]{\smash{\SetFigFont{5}{6.0}{\rmdefault}{\mddefault}{\updefault}{\color[rgb]{0,0,0}$+$}%
}}}
\put(2851,614){\makebox(0,0)[lb]{\smash{\SetFigFont{5}{6.0}{\rmdefault}{\mddefault}{\updefault}{\color[rgb]{0,0,0}$-$}%
}}}
\put(3226,314){\makebox(0,0)[lb]{\smash{\SetFigFont{5}{6.0}{\rmdefault}{\mddefault}{\updefault}{\color[rgb]{0,0,0}$R_6\ge\frac{1}{\sqrt{2}}$}%
}}}
\put(3226,-1336){\makebox(0,0)[lb]{\smash{\SetFigFont{5}{6.0}{\rmdefault}{\mddefault}{\updefault}{\color[rgb]{0,0,0}$R_6\ge\frac{1}{\sqrt{2}}$}%
}}}
\put(2401,-736){\makebox(0,0)[lb]{\smash{\SetFigFont{7}{8.4}{\rmdefault}{\mddefault}{\updefault}{\color[rgb]{0,0,0}$\Omega$}%
}}}
\put(1501,-1036){\makebox(0,0)[lb]{\smash{\SetFigFont{5}{6.0}{\rmdefault}{\mddefault}{\updefault}{\color[rgb]{0,0,0}$-$}%
}}}
\put(1501,614){\makebox(0,0)[lb]{\smash{\SetFigFont{5}{6.0}{\rmdefault}{\mddefault}{\updefault}{\color[rgb]{0,0,0}$+$}%
}}}
\put(976,-736){\makebox(0,0)[lb]{\smash{\SetFigFont{7}{8.4}{\rmdefault}{\mddefault}{\updefault}{\color[rgb]{0,0,0}$\Omega$}%
}}}
\put(3226,-136){\makebox(0,0)[lb]{\smash{\SetFigFont{5}{6.0}{\rmdefault}{\mddefault}{\updefault}{\color[rgb]{0,0,0}$R_6\le\frac{1}{\sqrt{2}}$}%
}}}
\put(3226,-1786){\makebox(0,0)[lb]{\smash{\SetFigFont{5}{6.0}{\rmdefault}{\mddefault}{\updefault}{\color[rgb]{0,0,0}$R_6\le\frac{1}{\sqrt{2}}$}%
}}}
\put(5476,314){\makebox(0,0)[lb]{\smash{\SetFigFont{5}{6.0}{\rmdefault}{\mddefault}{\updefault}{\color[rgb]{0,0,0}$R_5\ge\sqrt{2}$}%
}}}
\put(5476,-136){\makebox(0,0)[lb]{\smash{\SetFigFont{5}{6.0}{\rmdefault}{\mddefault}{\updefault}{\color[rgb]{0,0,0}$R_5\le\sqrt{2}$}%
}}}
\put(5476,-1336){\makebox(0,0)[lb]{\smash{\SetFigFont{5}{6.0}{\rmdefault}{\mddefault}{\updefault}{\color[rgb]{0,0,0}$R_5\ge\sqrt{2}$}%
}}}
\put(5476,-1786){\makebox(0,0)[lb]{\smash{\SetFigFont{5}{6.0}{\rmdefault}{\mddefault}{\updefault}{\color[rgb]{0,0,0}$R_5\le\sqrt{2}$}%
}}}
\put(1501,-2836){\makebox(0,0)[lb]{\smash{\SetFigFont{12}{14.4}{\rmdefault}{\mddefault}{\updefault}{\color[rgb]{0,0,0}{\bf (a)}}%
}}}
\put(4426,-2836){\makebox(0,0)[lb]{\smash{\SetFigFont{12}{14.4}{\rmdefault}{\mddefault}{\updefault}{\color[rgb]{0,0,0}{\bf (b)}}%
}}}
\put(6676,-2836){\makebox(0,0)[lb]{\smash{\SetFigFont{12}{14.4}{\rmdefault}{\mddefault}{\updefault}{\color[rgb]{0,0,0}{\bf (c)}}%
}}}
\put(6301,-211){\makebox(0,0)[lb]{\smash{\SetFigFont{9}{10.8}{\rmdefault}{\mddefault}{\updefault}{\color[rgb]{0,0,0}D}%
}}}
\put(6301,-1861){\makebox(0,0)[lb]{\smash{\SetFigFont{9}{10.8}{\rmdefault}{\mddefault}{\updefault}{\color[rgb]{0,0,0}D}%
}}}
\put(901, 89){\makebox(0,0)[lb]{\smash{\SetFigFont{12}{14.4}{\rmdefault}{\mddefault}{\updefault}{\color[rgb]{0,0,0}$+$}%
}}}
\put(2251, 14){\makebox(0,0)[lb]{\smash{\SetFigFont{12}{14.4}{\rmdefault}{\mddefault}{\updefault}{\color[rgb]{0,0,0}$-$}%
}}}
\put(2251,-1561){\makebox(0,0)[lb]{\smash{\SetFigFont{12}{14.4}{\rmdefault}{\mddefault}{\updefault}{\color[rgb]{0,0,0}$+$}%
}}}
\put(901,-1636){\makebox(0,0)[lb]{\smash{\SetFigFont{12}{14.4}{\rmdefault}{\mddefault}{\updefault}{\color[rgb]{0,0,0}$-$}%
}}}
\end{picture}
\end{center}
\caption{{\scriptsize 
The decay channels of $\Z_2$-charged branes are most easily seen as an $\Omega$ invariant
process in the $\I_4$ orbifold. The first line in the figure shows the standard decent of an 
$s=2$ D\db pair (a), via an $s=1$ \dh-brane (b), into an $s=0$ D\db pair. The second line is the
$\Omega$-image of this decay. Together, the diagrams show the decays between 
$\Z_2$-charged $s=2,1,0$-branes in the GP orientifold. 
{\bf (a)} A $\Z_2$-charged (5,2)-brane (called $h$ in the text) is an $\Omega$ invariant configuration 
of four fractional $(5,2)$-branes in the orbifold. The twisted \RR charges of each of the fractional
branes are shown as $\pm $ next to the fixed points denoted by crosses. The untwisted \RR charge is 
$\pm 1$ and shown in the middle of each brane. 
{\bf (b)} A $\Z_2$ charged (5,1)-brane is 
an $\Omega$ invariant configuration 
of two truncated $(5,1)$-branes in the orbifold. The twisted \RR charges of each of the \dh-branes are 
shown as $\pm $ next to the fixed points denoted by crosses. 
{\bf (c)} In the orbifold cover a stuck (5,0)-(anti-)brane is an $\Omega$-invariant pair of fractional 
$(5,0)$-(anti-)branes with opposite twisted \RR charges. Here we show a stuck brane at one fixed point with 
a stuck anti-brane at the other.}}
\label{fig5}
\end{figure}

In order to analyse the decay products of these torsion branes a useful 
way to think about them is as $\Omega$ invariant bound states of fractional or truncated branes 
in the $\I_4$ orbifold. Consider a (5,1)-brane stretching in the $x^5$ direction for example. 
In the orbifold this is an $\Omega$ invariant pair of \dh$(5,1)$-branes with opposite 
twisted \RR charges at both fixed points, as can be seen in Figure~\ref{fig5}~(b). Each of the 
$(5,1)$-branes is unstable for $R_5\ge\sqrt{2}$ and $R_i\le 1/\sqrt{2}$ $(i=6,7,8)$. For 
$R_5\ge\sqrt{2}$ each of the truncated branes decay into a brane-anti-brane pair of fractional 
$(5,0)$-branes at the two fixed points as is shown in figure~\ref{fig5}~(c). The decay products are 
$\Omega$ invariant as well and form a stuck $(5,0)$-brane at one fixed point and a stuck 
$(5,0)$-anti-brane at the other. At the fixed planes in the GP 
orientifold there will typically be $n_1,n_2$ tadpole-canceling 
(5,0)-branes, which will combine with the brane-anti-brane pair from the 
decay of the $(5,1)$-brane to give a BPS tadpole canceling configuration 
of $n_1+1,n_2-1$ (5,0)-branes. This process looks very much like the 
brane transfer of~\cite{BGH}. Each of the truncated (5,1)-branes also decays for  
$R_i\le 1/\sqrt{2}$ into a D\db$(5,2)$-pair as shown in Figure~\ref{fig5}~(a). 
Again the decay configuration is $\Omega$ 
invariant and forms a $\Z_2$ torsion charged $(5,2)$-brane in the orientifold.
By T-duality a similar argument can be made 
for the $s=3$ brane decaying into an $s=2$ brane as well as
into a brane-anti-brane pair of stuck $(5,4)$-branes. Thus by considering the branes in the orbifolded
cover we have related their torsion charges to one another.

\section{Open strings and Chan-Paton factors}\label{sec5}
\setcounter{equation}{0}

Many D-branes we have encountered throughout this paper can be thought of 
as bound states of branes from the $\I_4$ orbifold which are $\Omega$ 
invariant. These branes have non-trivial Chan-Paton factors on which the 
orientifold group as well as the GSO projection have to form, up to 
phases, a representation. Any such phases must be compensated for by 
opposite phases on the world-sheet fields. This was already encountered 
in~\cite{GP,Gproj}. In particular in~\cite{GP} 
it was argued that $\Omega^2=-1$ on world-sheet fields 
for open strings stretching between the D9- and D5-branes. To cancel this 
phase the representation of $\Omega$ on Chan-Paton factors of the D5-brane 
is anti-symmetric giving an $Sp$ instead of an $SO$ gauge group on the D5-branes. 
In this section 
we discuss several cases where  such phases enter the representation 
of the orientifold group and GSO projection. We first consider the $\Z_2$ 
D7-brane of Type I which is an $\Omega$ invariant pair of Type IIB 
D7-branes of opposite \RR charge. We show that $\Omega$ and $(-1)^F$ form 
a projective representation of $\Z_2\times\Z_2$ on the Chan-Paton 
factors, and discuss how this phase is compensated for. In the second part of this
section we show how this discussion generalises to certain branes in the GP model.

\subsection{The D7-brane of Type I and projective representations of 
$\Omega\times(-1)^F$}

The D7-brane of Type I is an $\Omega$ invariant bound state of D7-branes 
with opposite bulk \RR charge.\footnote{A similar analysis applies to the 
D-instanton of Type I.} The Chan-Paton factors form a $2\times 2$ matrix
\be
\lambda=\left(\begin{array}{cc} a&b\\c&d\end{array}\right)\,,
\ee
where $a$, $d$ is the open string with both endpoints on the brane, 
anti-brane, respectively and $b,c$ correspond to the open strings that 
stretch between the brane and the anti-brane. The GSO projection on $a$ 
and $d$ is $(1+(-1)^F)/2$, while on $b$ and $c$ it is 
$(1-(-1)^F)/2$. This can be compactly expressed as
\be
\gamma_{(-1)^F}=\left(\begin{array}{cc} 1&0\\0&-1\end{array}\right)
\ee
and
\be
(-1)^F:(\lambda)\rightarrow \gamma_{(-1)^F}\lambda\gamma^{-1}_{(-1)^F}=
\left(\begin{array}{cc} a&-b\\-c&d\end{array}\right)\,.
\ee
We want to find the representation of $\Omega$ on the Chan-Paton factors 
which together with $\gamma_{(-1)^F}$ should, up to $U(1)$ phases form a 
$\Z_2\times\Z_2$ representation. These $U(1)$ phases will have to be 
cancelled by opposite phases on the world-sheet fields such that on the 
full open string Hilbert space we get a proper
representation.\footnote{This is needed to show that $(1+\Omega)/2$ and 
$(1+(-1)^F)/2$ are genuine projection operators.} Thus
\be
\gamma_{(-1)^F}\gamma_\Omega=e^{i\theta}\gamma_\Omega\gamma_{(-1)^F}\,,
\ee
for some $\theta$. As a result $\gamma_\Omega$ has to be either
\be
\left(\begin{array}{cc}*&0\\0&*\end{array}\right)\,,
\ee
or
\be
\left(\begin{array}{cc}0&* \\ \mbox{$*$}& 0\end{array}\right)\,.
\label{gammaform}
\ee
The one-loop partition function for an open string ending on the D7-brane 
was computed in~\cite{Frau} where one finds that the full M\"obius strip amplitude with 
and without $(-1)^F$ inserted come with the same sign. On the world-sheet 
fields the two traces come with opposite signs due to the action of 
$(-1)^F$ on the vacuum. The Chan-Paton traces must compensate
\be 
\tr(\gamma^T_\Omega\gamma^{-1}_\Omega)=
-\tr(\gamma^T_{\Omega(-1)^F}\gamma^{-1}_{\Omega(-1)^F})\,,
\ee
and so $\gamma_\Omega$ is off-diagonal~(\ref{gammaform}).\footnote{This off-diagonal form 
of $\Omega$ is consistent with the geometric picture~\cite{WittK} where it is argued 
that $\Omega$ interchanges the $a,d$ open strings and leaves invariant (up to phase) the $b,c$ ones.}

In order to see that this forms a projective representation rather than a 
proper representation we consider an open string stretching between the 
D7-brane and a brane on which $\Omega$ and $(-1)^F$ form a proper 
representation on the Chan-Paton sector. For simplicity we consider a 
single D-string which has trivial Chan-Paton factors. For a D7-D1 
(D1-D7) string the Chan-Paton factors form a two dimensional column 
(row) vector. $\Omega$ exchanges a D7-D1 string with a D1-D7 string and 
vice-versa. It is then straightforward to check that for a D7-D1 string
\be
\lambda^T\gamma^{-1}_\Omega\gamma^{-1}_{(-1)^F}=-
(\gamma_{(-1)^F}\lambda)^T\gamma^{-1}_\Omega\,,
\ee
where $\lambda$ is the Chan-Paton matrix for the D7-D1 string; a similar 
result holds for the D1-D7 string. In other words on Chan-Paton factors 
of open strings stretching between the D7- and D1- branes we have
\be
\Omega(-1)^F=-(-1)^F\Omega\,.
\label{anticomm}
\ee
On the full open string Hilbert space $\Omega$ and $(-1)^F$ commute and 
form a proper representation. At first it would appear that $\Omega$ and 
$(-1)^F$ should anti-commute on the world-sheet fields; in other words 
that $\Omega$ should be fermionic! On the other hand $\Omega$ maps the 17 
sector to the 71 sector and so $(-1)^F$ on the left-hand side of 
equation~(\ref{anticomm}) acts on a different sector to $(-1)^F$ on 
the right-hand side. By having a relative minus sign between the two 
sectors' $(-1)^F$'s, $\Omega$ and $(-1)^F$ will commute on the full 
Hilbert space. Since the 17 and 71 strings have $4k+2$ zero modes the 
action of $(-1)^F$ on them is only well defined up to a sign and so 
$(-1)^F$ can indeed be defined with an extra minus sign in the 17 sector 
relative to the 71 sector.

\subsection{Chan-Paton factors and projective representations in the GP 
orientifold}

The results of the last subsection easily generalise to the branes we 
have been constructing in the GP orientifold. In particular for the 
(5,2)-branes with twisted \NSNS couplings $\Omega\times (-1)^F$ also form 
a projective representation on Chan-Paton factors, while $I_4$ acts 
trivially. Similarly, since the truncated $(-1,0)$- and $(-1,4)$-branes 
are $\Omega$ invariant fractional brane-anti-brane pairs they too have 
non-trivial Chan-Paton factors. It is easy to see that on these both 
$(-1)^F$ and $\I_4$ act as
\be
\left(\begin{array}{cc} 1&0\\0&-1\end{array}\right)\,.
\ee
From the M\"obius strip diagrams in Appendix~\ref{appc}, as in the 
previous subsection, one finds that $\Omega$ has an off-diagonal form. As 
above this gives a projective representation on the CP factors. The phases 
encountered here have to be cancelled by opposite phases on the 
world-sheet fields if the orientifold $\times$ GSO group is to form a 
proper representation on the full open string Hilbert space. As in the 
previous subsection the only way that this can be done for a bosonic 
$\Omega$ is by having $(-1)^F$ and $\I_4$ acting on $pp'$ strings with an 
extra relative minus sign as compared with the $p'p$ strings. Here $p$ is 
the brane with a projective representation and $p'$ is a brane with a 
proper representation of the orientifold $\times$ GSO group. In all cases 
there are $4k+2$ fermionic zero modes and so the action of $(-1)^F$ and 
$\I_4$ is again only well defined up to an overall sign.

\section{Some comments on K-theory}\label{sec6}
\setcounter{equation}{0}

D-branes can be viewed as bundles over sub-manifolds of spacetime. 
Physically one is only interested in isomorphism classes of these and, 
since brane-anti-brane annihilation is allowed,
the classes of objects we are 
interested are best described by K-groups. Depending on the exact nature 
of the allowed bundles one may study many different types of K-groups: in 
Type II theories complex K-groups, $K^*$, in Type I orthogonal K-groups, 
$KO^*$ and in orbifolds equivariant K-groups $K_G^*$~\cite{WittK} (see also~\cite{garcia}). 
It would be natural 
to conjecture that Type I orbifolds would  be described by equivariant, 
orthogonal K-groups, $KO_G^*$. Yet we have seen that if the orientifold 
group $\Omega\times G$ admits projective representations (as can be seen 
by computing $H^2(\Z_2\times G,U(1))$) there is a choice in picking the 
action of $\Omega$ on certain twisted sectors. In the case studied 
explicitly here there are two inequivalent theories with either a tensor 
or hypermultiplet in each twisted sector. As a result the D-brane 
spectrum is very different in the two theories, and so one expects two 
different K-groups. One of these will have to be $KO_{\Z_2}$ while the 
other will be a twisted version, $KO^{[c]}_{\Z_2}$ of this group in which 
the non-trivial two-cocycle $c$ of $H^2(\Z_2\times \Z_2,U(1))=\Z_2$ is 
used to twist. Such a group, to the best of our knowledge, has 
unfortunately not been considered in the mathematical literature. 

It is immediate that
\be
KO_{\Z_2}(\mbox{pt})=\Z\oplus\Z
\ee
and we may conclude that there should be two charges associated with the 
$(5,4)$-brane. This happens in the DPBZ model where twisted \RR six-form 
potentials exist and the $(5,4)$-branes in the non-compact theory carry 
an untwisted as well as a twisted \RR charge giving $\Z\oplus\Z$. In the 
GP model on the other hand the $(5,4)$-branes are only charged under the 
untwisted \RR charge. Clearly then it is the DPBZ model which should be 
described by $KO_{\Z_2}$ while the GP model should be described by 
$KO^{[c]}_{\Z_2}$ for which we expect
\be
KO^{[c]}_{\Z_2}=\Z\,.
\ee
The presence of twisted K-theories is related to the presence of a 
\NSNS flux~\cite{WittK}. Such a flux 
in orientifold theories changes the charge of an orientifold plane~\cite{kir}.
In our case the difference between GP and DPBZ comes precisely from the two
types of O5-planes present in the theories.

The relationship between $KO_{\Z_2}$ and $KO^{[c]}_{\Z_2}$ is expected to be
analogous to the relationship between $K_{\Z_2\times\Z_2}$ and $K^{[c]}_{\Z_2\times\Z_2}$.
These last two K-theories describe D-branes in Type II on the $\Z_2\times\Z_2$ 
orbifold~\footnote{For details of various aspects of these orbifolds 
see~\cite{vafadt,z2torsion,WittK,SCY,Gproj}.} with the twisted theory corresponding to the orbifold with discrete 
torsion. In $K^{[c]}_{\Z_2\times\Z_2}$ the two generators of $\Z_2\times\Z_2$ 
anticommute on the fibers, since they form a projective representation of the orbifold group.
For $KO^{[c]}_{\Z_2}$ it is similarily expected that complex conjugation and the
generator of $\Z_2$ should anticommute.

In the above we have restricted ourselves to a discussion of the orientifold group 
$\Omega\times\I_4$. It is however immediate that for any orientifold group
$\Omega\times G$, where $G$ has order 2 elements, the action of $\Omega$ on the respective
twisted sectors will be ambigous.\footnote{Similar results should hold for 
orientifolds in which $\Omega$ is always combined with some geometric involution, where
the description in terms of KR rather than KO and its twisted analogues is more natural.}
 As a result there should be twisted KO-theories
corresponding to the various $\Omega$ actions on the twisted sectors. For example the 
for the $\Omega\times\Z_3$ orientifold $\Omega$ has a unique action on the twisted sectors
and so we expect that $KO_{\Z_3}$ should describe D-branes in this background. However, for
the $\Omega\times\Z_2\times\Z_2$ orientifold, as well as $KO_{\Z_2\times\Z_2}$ we expect three
twisted K-theories $KO^{c_i}_{\Z_2\times\Z_2}$, $i=1,2,3$ corresponding to the 
four different choices of $\Omega$ actions on the twisted sectors.

\section{Branes on the DPBZ orientifold}\label{sec7}
\setcounter{equation}{0}

In this section we extend our analysis of the GP model to the DPBZ model. 
Since the computations are rather similar to the ones discussed above for the GP model we limit ourselves
here to a summary of the stable branes in the 
DPBZ model. The BPS fractional branes are the $(1,0)$- and $(1,4)$-branes as well as the tadpole 
canceling $(5,0)$- and $(5,4)$-branes.\footnote{In order to
cancel the \RR twisted sector tadpole, the $(5,0)$- and $(5,4)$-branes have opposite twisted \RR charges.}
The BPS stuck branes are the $(-1,2)$ and $(3,2)$-branes.

The integrally charged non-BPS truncated branes exist for $r=1,5$ and $s=1,2,3$. 
There are again two kinds of torsion charged non-BPS branes. The branes that couple to twisted \NSNS 
sectors have $r=-1,0$ and $s=0$ or $r=3,4$ and $s=4$. However, unlike the GP model, these branes 
{\em do not} develop open string tachyons on their world-volumes for any values of the compactification 
radii (\cf equation~(\ref{condtortor})). 
The open strings stretching between the $r=3,4$ and $s=4$ torsion branes 
and the tadpole canceling $(5,4)$-branes have tachyonic ground-states as in the GP model. On the other
hand the $(-1,0)$- and $(0,0)$-branes 
({\em i.e.} the D-particle and D-instanton) are stable in the presence of the tadpole canceling branes, with the
D-particle carrying a spinor charge of one of the $SO(8)$'s. 
As in the GP model there are also $\Z_2$-charged branes coupling only to the untwisted \NSNS sector.  
They exist for $r=-1,0$ and $s=0,1$ as well as for $r=3,4$ and $s=3,4$. Most of these have the usual 
stability conditions~(\ref{full}). Only the $(0,1)$- is stable for $R_i\le\sqrt{2}$ 
($i=5,\dots,4+s$) and the $(4,3)$-brane is stable for $R_8\ge 1/\sqrt{2}$ with the branes stretching
along $x^5,\dots,x^{4+s}$ in the internal torus.

\section{Conclusions and Outlook}\label{sec8}
\setcounter{equation}{0}

In this paper we have constructed all stable BPS and non-BPS D-branes in 
the GP and DPBZ orientifolds. As well as the expected BPS branes, 
these include both integral and torsion charged non-BPS D-branes.
We have found that integrally charged non-BPS D-branes come in two forms: either as
truncated D-branes similar to the ones encountered in orbifolds~\cite{GS}, or as 
$\Omega$ invariant BPS-anti-BPS pairs of branes. 
We have also found a rich spectrum of stable torsion charged D-branes. In 
particular we have found a new class of torsion branes which 
couple to twisted and untwisted \NSNS sectors. The orientifold theories also
have a large number of torsion branes coupling only to the untwisted \NSNS sector. 
These resemble the Type I-like theory torsion branes~\cite{BGH}.

We have identified the stability conditions and decay products of the non-BPS 
D-branes. Due to the $\Omega$ projection the truncated branes have stability regions
which are quite different from the orbifold stability regions. The $\Omega$ projection
also has a significant effect on the torsion branes' stability regions. In the GP
model, the stability of the torsion branes coupling to the twisted \NSNS sector resembles 
those encountered in~\cite{SCY,MajSen}, while in the DPBZ model these branes are in fact 
always stable. 

Many of the branes in orientifolds have non-trivial 
Chan-Paton factors and we have found that the Orientifold $\times$ GSO 
group can form projective representations on these. In fact even for the 
$\Z_2$ D$7$-brane and D-instanton of Type I $\Omega\times (-1)^F$, form a 
projective representation on the Chan-Paton factors. 

We have discussed the role of K-theory in Type I orbifolds and have found that 
the equivariant orthogonal K-group $KO_G$ does {{\em not} always give the 
D-brane spectrum. In particular, if $G$ has order two elements 
new twisted KO-groups describe the D-brane spectrum. This twisting should be similar 
to the twisted K-groups used in studying orbifolds of Type II theories 
with discrete torsion.

It would be interesting to understand better these twisted K-theories, as 
well as have a better understanding of the charges associated to branes 
coupling to twisted \NSNS sectors. Further the decay products of the 
GP-model $(5,2)$-brane with twisted \NSNS couplings have not been identified. Since the brane is 
really a pair of branes in the cover it may be that the analysis 
of~\cite{MajSen} is applicable. We hope to return to these issues in the 
future.

\section*{Acknowledgments}

We are grateful to G. Arutyunov, C. Angelantonj, V. Braun, S. Fredenhagen, D. 
Ghoshal, A. Keurentjes, D. Husemoller, J. Polchinski, R. Rabadan, E. Scheidegger, V. 
Schomerus and B. Totaro for useful conversations. We would especially 
like to thank M.R. Gaberdiel, F. Quevedo and S. Theisen for their continued interest, 
support and valuable comments throughout this project. We are grateful to 
the organisers of the M-theory Cosmology conference and the TMR Corfu meeting 
for providing a stimulating environment during the completion of this work.
N.Q. is funded by GIF and DAAD.

\appendix

\section{Coherent states in orientifolds}\label{appa}
\setcounter{equation}{0}

In this appendix we construct coherent states corresponding to D-branes 
and O-planes in the GP orientifold. The coherent states have to be 
invariant under the closed-string GSO projection as well as under 
$\Omega$ and $\I_4$. $\Omega$ does not introduce any closed string 
twisted sectors, hence up to normalisation, the boundary states 
corresponding to D-branes constructed in~\cite{GS} need simply be 
projected by $\ms{$\half$}(1+\Omega)$. This is discussed in 
Appendix~\ref{appb}. The normalisation of the boundary states does 
change. Firstly, there will be an extra factor of $1/2$ for ${\cal N}^2$ 
coming from the orientation projection on the open strings partition 
function. Secondly, ${\cal N}^2$ will acquire an extra factor of 
$(\Tr(\gamma(1))^2$, coming from the Chan-Paton degrees of freedom. The 
boundary states' normalisation coefficients are given in Appendices~\ref{appc} and~\ref{appd}.

The crosscaps corresponding to O-planes can be constructed by perusal of 
the Klein-bottle amplitudes ${\cal K}$ in the tree channel. In the loop channel
\be
{\cal K}=\int_0^\infty\frac{\rmd 
t}{2t}\Tr_{\ms{\NSNS,\RR}}^{\ms{U+T}}\left(\frac{\Omega}{2}
\frac{1+(-1)^F}{2}\frac{1+\I_4}{2}e^{-2\pi tH_c}\right)\,,
\ee
where the trace is over all untwisted and twisted bosonic sectors, $H_c$ 
is the closed string Hamiltonian, and the GSO projection is only taken in 
the left moving sector as the states have to be left-right symmetric. 
Evaluating we get
\ba
{\cal K}&=&\frac{V_6}{(2\pi)^6}\frac{1}{8}\int\frac{\rmd t}{2t}t^{-3}
\frac{f_3^8({\tilde q}^2)-f_4^8({\tilde q}^2)-f_2^8({\tilde 
q}^2)}{f_1^8({\tilde q}^2)}
\prod_{i=5}^8\left(\sum_{n_i\in\Z}e^{-\pi 
t(n_i/R_i)^2}+\sum_{w_i\in\Z}e^{-\pi
t(w_iR_i)^2}\right)
\nonumber \\&&+\,\,
16\frac{V_6}{(2\pi)^6}\frac{1}{4}\int\frac{\rmd t}{2t}t^{-3}
\frac{f_3^4({\tilde q}^2)f_2^4({\tilde q}^2)-f_2^4({\tilde 
q}^2)f_3^4({\tilde q}^2)}
{f_1^4({\tilde q}^2)f_4^4({\tilde q}^2)}\nonumber \\
&=&\frac{1}{2}\frac{V_6}{(2\pi)^6}2^4\int\rmd l
\frac{f_3^8(q)-f_2^8(q)-f_4^8(q)}{f_1^8(q)}
\prod_{i=5}^8\left(R_i\sum_{w_i\in 2\Z}e^{-\pi t(w_iR_i)^2}+\frac{1}{R_i}
\sum_{n_i\in 2\Z}e^{-\pi t(n_i/R_i)^2}\right)
\nonumber \\&&+\,\,
32\frac{V_6}{(2\pi)^6}\int\rmd l
\frac{f_3^4(q)f_4^4(q)-f_4^4(q)f_3^4(q)}{f_1^4(q)f_2^4(q)}\,,\label{kb}
\ea
where ${\tilde q}=e^{-\pi t}$, the $f_i$ are defined in~\cite{GP}, 
$l=1/(4t)$ is the tree channel modular parameter~\cite{GP}, $q=e^{-2\pi 
l}$, 
$V_6$ is the (infinite) volume of the six non-compact space-time 
directions, and $R_i$ are the radii of $T^4$. The two parts of the first 
integral correspond to $1$ and $\I_4$ insertions in the untwisted sector 
trace, respectively. The second integral comes from the trace over the sixteen twisted 
sector states. In the tree channel the two terms of the first integral 
correspond to the exchange between two O9- or two O5-planes, respectively, 
while the second integral corresponds to the O9-O5 interaction. 

In non-compact space-time one defines in each (bosonic) sector of the 
theory the crosscap state 
\be\label{boundary1}
\ket{C(r,s),k,\eta}=\exp\left(\sum_{l>0}^\infty 
\left[\frac{(-1)^l}{l}\alpha_{-l}^\mu 
S_{\mu\nu}\tilde{\alpha}_{-l}^\nu\right]
+i\eta\sum_{m>0}^\infty (-1)^m\left[\psi_{-m}^\mu
S_{\mu\nu}\tilde{\psi}_{-m}^\nu\right]\right)\ket{C(r,s),k,\eta}^{(0)}\,,
\label{bdr}
\ee
where, depending on the sector, $l$ and $m$ are integer or half-integer, 
$\eta=\pm 1$,  and $k$ denotes the momentum of the ground state.  
The matrix $S$ is diagonal and 
has (in Euclidean space-time) entries equal to $-1\,,1$ for Neumann and 
Dirichlet boundary conditions, respectively. The zero-mode part of the 
crosscap state is the same as that of a boundary state, and we refer the 
reader to Appendix~\ref{appb} and~\cite{GS} for a discussion of these. 

O-planes are localised in transverse space, so we Fourier transform the 
above crosscap state
\be
\label{localch3}
\ket{C(r,s),y,\eta}=\int\left(\prod_{\mu 
\;\mbox{\scriptsize{transverse}}} 
dk^\mu e^{ik^\mu y_\mu}\right) \ket{C(r,s),k,\eta}\,,
\ee
$y$ denotes the location of the O-plane. Compactifying some directions on 
circles modifies the zero mode part of the crosscap state. In particular, 
the momentum integrals become sums over $2n_i/R_i$ ($n_i\in\N$), and in 
compact Neumann directions the ground state becomes a sum over windings $2w_i 
R_i$ ($w_i\in\N$). The momenta and windings are even so as to match with 
the momentum and winding sums in the tree channel of equation~(\ref{kb}). 

As in the case of D-branes, closed string GSO invariance combines the 
$\eta=+,-$ crosscaps 
in each sector to give one state per closed string sector
\ba
\left|C(r,s)\right>_{\mbox{\scriptsize\NSNS}}&=&
\frac{1}{2}\Bigl(\left|C(r,s),+\right>_{\mbox{\scriptsize\NSNS}}-
\left|C(r,s),-\right>_{\mbox{\scriptsize\NSNS}}\Big)\,,\\
\left|C(r,s)\right>_{\mbox{\scriptsize\RR}}&=&
\frac{4i}{2}\Bigl(\left|C(r,s),+\right>_{\mbox{\scriptsize\RR}}+
\left|C(r,s),-\right>_{\mbox{\scriptsize\RR}}\Bigr)\,.
\ea
From the tree channel expression in equation~(\ref{kb}) one can read off 
the form as well as normalisations of the GP O9- and O5-planes
\ba
\ket{O(5,4)} & = & {\cal N}_9
\left(\left|C(5,4)\right>_{\mbox{\scriptsize\NSNS}} 
+\left|C(5,4)\right>_{\mbox{\scriptsize\RR}} \right) 
\nonumber \\
\ket{O(5,0),{\bf y}} & = & {\cal N}_5
\left(\left|C(5,0),{\bf y}\right>_{\mbox{\scriptsize\NSNS}} 
+\left|C(5,0),{\bf y}\right>_{\mbox{\scriptsize\RR}} \right)\,,
\ea
where ${\bf y}=(y_5,\dots,y_8)$ with $y_i=0,\pi R_i$ fixing the location 
of the O5-plane to one of the sixteen fixed points of $T^4/\Z_2$. The 
normalisation constants are
negative and given by\footnote{The fact that they are negative is to be 
expected; after all the O9 and O5-planes have negative tension and \RR 
charge. The sign of the normalisation constants follows from the M\"obius 
diagram.}
\be
{\cal N}_{\ms{C}9}^2=\frac{V_6}{(2\pi)^6}8\prod_{i=5}^8R_i\,,\qquad
{\cal N}_{\ms{C}5}^2=\frac{V_6}{(2\pi)^6}\frac{1}{32}\prod_{i=5}^8 
\frac{1}{R_i}\,.
\ee
Note that, due to the sum over {\em even} momenta, each of the $16^2$ 
O5-plane diagrams is the same. With the above definitions one may easily 
check that
\ba
{\cal K}=\int_0^\infty\rmd 
l\left(\bra{C(5,4)}+\sum_{i=1}^{16}\bra{C(5,0),{\bf y}_i}\right)
e^{-lH_c}\left(\ket{C(5,4)}+\sum_{i=1}^{16}\ket{C(5,0),{\bf y}_i}\right)\,,
\ea
with ${\bf y}_i$ labeling the position of the sixteen fixed points.

\section{$\Omega$ invariance}\label{appb}
\setcounter{equation}{0}

In this appendix we study the $\Omega$ invariance of boundary and 
crosscap states. Defining the action of $\Omega$ on closed string states 
as in~\cite{GP}
\be
\Omega\alpha_r\Omega^{-1}={\tilde\alpha}_r\,,\qquad
\Omega\psi_r\Omega^{-1}={\tilde\psi}_r\,,\qquad
\Omega{\tilde\psi}_r\Omega^{-1}=\psi_r\,,
\ee
it is easy to see that the non-zero mode part of the boundary and 
crosscap states is $\Omega$ invariant. As in the case of 
orbifolds~\cite{GS} constraints on $r$ and $s$ come from the analysis of 
the fermionic zero-modes. The various boundary states have to couple 
consistently to closed string states. For example in Type IIB the \RR 
sector boundary state has to couple to even \RR potentials. This in 
effect fixes the zero-mode part of the boundary state and is consistent 
with GSO-invariance. Let us see how this happens for $\Omega$. In the 
untwisted \RR sector the $\Omega$ projection is represented on the 
zero-modes by\footnote{For simplicity we work in the light-cone gauge here.}
\be
\Omega_{\ms{\RR}}=\kappa_{\ms{\RR}}\prod_{i=1}^8\frac{1-2\psi^i_0{\tilde\psi}^i_0}{\sqrt{2}}\,.
\ee
This satisfies the relation $\Omega^2=(-1)^{F+{\tilde F}}$, with $(-1)^F$ 
and $(-1)^{{\tilde F}}$ defined as in~\cite{BG} for 
$\kappa^2_{\ms{\RR}}=1$. From this one obtains
\be
\Omega\ket{B(r,s),\pm}^{(0)}_{\ms{\RR}}=\kappa_{\ms{\RR}}i^{11-p}
\ket{B(r,s),\pm}^{(0)}_{\ms{\RR}}\,.
\ee
Since we expect D$(r,s)$-branes with $p=r+s=1,5,9$ to 
survive\footnote{This must be the case as the \RR potentials present in 
Type I are $\C2$ and $C^{(10)}$.} we pick $\kappa_{\ms{RR}}=-1$.
The crosscap states' zero-mode part is the same as that of a boundary 
state, so the above holds for it as well. In the twisted \RR sector 
$\Omega$ is defined as above with $i=1,\dots,4$ and a different constant 
$\kappa_{\ms{\RRT}}$ (which also squares to one). When acting on the 
boundary state
\be
\Omega\ket{B(r,s),\pm}^{(0)}_{\ms{\RRT}}=\kappa_{\ms{\RRT}}i^{5-r}
\ket{B(r,s),\pm}^{(0)}_{\ms{\RRT}}\,.
\ee

In the GP orientifold the \RRT 6-form and 2-form are removed by $\Omega$. 
This fixes $\kappa_{\ms{\RRT}}=-1$. Finally, in the twisted \NSNS sector 
in the definition of $\Omega$ $i=5,\dots,8$, and we have
\be
\Omega\ket{B(r,s),\pm}^{(0)}_{\ms{\NSNST}}=\kappa_{\ms{\NSNST}}i^{6-s}
\ket{B(r,s),\pm}^{(0)}_{\ms{\NSNST}}\,.
\ee
Comparing to the massless spectrum fixes $\kappa_{\ms{\NSNST}}=1$. We 
have fixed $\kappa$ in the above such that the boundary states couple to 
closed string states which are present in the theory. As a result we 
expect that the action of $\Omega$ defined in this way corresponds to the 
one used in~\cite{GP}. In summary the following boundary states are 
$\Omega$ invariant
\be
\begin{array}{ll}
\ket{B(r,s)}_{\mbox{\scriptsize\NSNS}} & \hbox{for all $(r,s)$}\,, \\
\ket{B(r,s)}_{\mbox{\scriptsize\RR}} & r+s=1,5,9\,, \\
\ket{B(r,s)}_{\mbox{\scriptsize\NSNS,T}} & s=2\,, \\
\ket{B(r,s)}_{\mbox{\scriptsize\RR,T}} & r=-1,3\,,
\end{array}
\ee
in agreement with~\cite{EP}.

\section{Normalisations and partition functions of integrally charged branes}\label{appc}
\setcounter{equation}{0}

In this appendix we compute the one-loop partition function for open 
strings that end on truncated branes so as to verify that they are 
tachyon free for certain ranges of the compactification radii. By 
comparing with the open string one-loop partition functions we also 
obtain the normalisation of these boundary states corresponding to the branes.
The partition functions are computed in the tree channel and modular 
transformed to the loop channel {\em i.e.} the open string partition function. 
The relevant diagrams have the topologies of the annulus and M\"obius strip
\ba
{\cal A}&=&\int_0^\infty\rmd 
l\bra{\mbox{\dh}(r,s)}e^{-lH_{\ms{c}}}\ket{\mbox{\dh}(r,s)}\,,\\
{\cal M}_9&=&\int_0^\infty\rmd 
l\bra{C(5,4)}e^{-lH_{\ms{c}}}\ket{\mbox{\dh}(r,s)}+
\bra{\mbox{\dh}(r,s)}e^{-lH_{\ms{c}}}\ket{C(5,4)}\,,\\
{\cal M}_5&=&\int_0^\infty\rmd l\sum_{i=1}^{16}\left(\bra{C(5,0),{\bf y}_i}e^{
lH_{\ms{c}}}\ket{\mbox{\dh}(r,s)}+
\bra{\mbox{\dh}(r,s)}e^{-lH_{\ms{c}}}\ket{C(5,0),{\bf y}_i}\right)\,,
\ea
with $\ket{\mbox{\dh}(r,s)}$ defined in equation~(\ref{truncbrane}). Evaluating
\ba
{\cal A}&=&\half{\cal N}^2_{(r,s),U}\int_0^\infty\rmd l 
l^{(r-5)/2}\frac{f^8_3(q)
f^8_4(q)}{f_1^8(q)}
\prod_{i=5}^{4+s}\sum_{w_i\in\Z}e^{-\pi 
l(w_iR_i)^2}\prod_{j=5+s}^8\sum_{n_j\in\Z}e^{-\pi
l(n_j/R_j)^2}
\nonumber  \\
&&-\;\;2^{s-1}{\cal N}^2_{(r,s),T}\int_0^\infty\rmd l l^{(r-5)/2}
\frac{f^4_2(q)f^4_3(q)}{f_1^4(q)f_4^4(q)}\,,
\\
{\cal M}_9&=&{\cal N}_{(r,s),U}{\cal N}_{\ms{C}9}
\int_0^\infty\rmd l 
\frac{f^{r+s-1}_3(iq)f^{9-r-s}_4(iq)-f^{r+s-1}_4(iq)f^{9-r-s}_3(iq)}
{f^{r+s-1}_1(iq)f^{9-r-s}_2(iq)2^{(r+s-9)/2}}
\prod_{i=5}^{4+s}\sum_{w_i\in 2\Z}e^{-\pi l(w_iR_i)^2}\,,\\
{\cal M}_5&=&16{\cal N}_{(r,s),U}{\cal N}_{\ms{C}5}
\int_0^\infty\rmd l 
\frac{f^{3+r-s}_3(iq)f^{5-r+s}_4(iq)-f^{3+r-s}_4(iq)f^{5-r+s}_3(iq)}
{f^{3+r-s}_1(iq)f^{5-r+s}_2(iq)2^{(r-s-5)/2}}\!\!
\prod_{j=5+s}^8\sum_{n_j\in 2\Z}e^{-\pi l(n_j/R_j)^2}\,,\nonumber \\
\ea
modular transforming to the open channel these become\footnote{The 
annulus and M\"obius strip 
modular transformations are respectively $t=1/(2l)$ and $t=1/(8l)$.}
\ba
{\cal A}&=&{\cal N}^2_{(r,s),U}2^4\frac{\prod_{j=5+s}^8 
R_j}{\prod_{i=5}^{4+s} R_i}
\int_0^\infty\frac{\rmd t}{2t} (2t)^{-(r+1)/2}\nonumber 
\\&&\qquad\qquad\qquad\qquad\times\;\;
\frac{f^8_3(\tq)-f^8_2(\tq)}{f_1^8(\tq)}
\prod_{i=5}^{4+s}\sum_{n_i\in\Z}e^{-2\pi 
t(n_i/R_i)^2}\prod_{j=5+s}^8\sum_{w_j\in\Z}e^{-2\pi
t(w_jR_j)^2}
\nonumber  \\
&&-\;\;{\cal N}^2_{(r,s),T}2^{s+2}\int_0^\infty\frac{\rmd t}{2t} 
(2t)^{-(r+1)/2}
\frac{f^4_4(\tq)f^4_3(\tq)}{f_1^4(\tq)f_2^4(\tq)}\,,
\\
{\cal M}_9&=&\frac{{\cal N}_{(r,s),U}{\cal 
N}_{\ms{C}9}}{2\prod_{i=5}^{4+s} R_i}
\int_0^\infty\frac{\rmd t}{2t}(2t)^{-(r+1)/2} \prod_{i=5}^{4+s}\sum_{n_i\in\Z}e^{-2\pi t(n_i/R_i)^2}
\nonumber \\&&\times\;\;
\frac{e^{i\frac{\pi}{4}(r+s-5)}f^{r+s-1}_4(i\tq)f^{9-r-s}_3(i\tq)-e^{-i\frac{\pi}{4}(r+s-5)}
f^{r+s-1}_3(i\tq)f^{9-r-s}_4(i\tq)}{f^{r+s-1}_1(i\tq)f^{9-r-s}_2(i\tq)2^{(r+s-9)/2}}
\,,\nonumber \\ \\
{\cal M}_5&=&8{\cal N}_{(r,s),U}{\cal N}_{\ms{C}5}\prod_{j=5+s}^8 R_j
\int_0^\infty\frac{\rmd t}{2t}(2t)^{-(r+1)/2}\prod_{j=5+s}^8\sum_{w_j\in\Z}e^{-2\pi t(w_jR_j)^2}
\nonumber \\&&\times\;\;
\frac{e^{i\frac{\pi}{4}(r-s-1)}f^{3+r-s}_4(i\tq)f^{5-r+s}_3(i\tq)-e^{-i\frac{\pi}{4}(r-s-1)}
f^{3+r-s}_3(i\tq)f^{5-r+s}_4(i\tq)}{f^{3+r-s}_1(i\tq)f^{5-r+s}_2(i\tq)2^{(r-s-5)/2}}
\,.\nonumber \\
\ea
With the normalisations in Table~\ref{tb1} as well as the normalisations of crosscaps in 
Appendix~\ref{appa} it is easy to see that the open string groundstate tachyon 
cancels for $s=0\,,4$ and all $r$ and 
for $r=-1,3$ and all $s$. As discussed in section~\ref{sec3} \dh-branes 
exist only for $r=-1,3$ and $s\neq 2$.
Note that the tachyon ground-state cancellation is independent of the 
overall normalisation of the \dh-brane boundary states confirming that 
the \dh-branes are indeed integrally charged. 

The normalisation of the boundary states can be obtained most easily by 
comparing with the one-loop partition functions computed in the loop 
channel. Particular attention needs to be paid to the Chan-Paton factors 
which will affect the normalisations - for example ${\cal N}_U$ will be 
proportional to the trace of the identity operator, $\tr(1)$, on the 
Chan-Paton states. We summarise the normalisations of truncated as well 
as fractional and stuck branes in Table~\ref{tb1}.

\begin{table}[hbt] \begin{center} \begin{tabular}{|c|c|c|c|c|c|}\hline
$(r,s)$ & (-1,0), (1,0), (-1,4), & (-1,1), (-1,3) & (3,0), (3,4) & (3,1), (3,3) & 
(-1,2) \\ 
 & (3,2), (1,4), (5,0), (5,4) & & & & \\ 
\hline
&&&&&\\
$n^2$ & $\frac{1}{32}$ & $\frac{1}{64}$ & $\frac{1}{8}$ & $\frac{1}{16}$ 
& $\frac{1}{128}$ \\
&&&&&\\
\hline 
\end{tabular}
\caption{Normalisations of boundary states for minimally, integrally 
charged $(r,s)$-branes. $n$ is related to the normalisations via ${\cal 
N}^2_{U,(r,s)}=\frac{V_{r+1}}{(2\pi)^{r+1}}n^2
\frac{\prod_{i=1}^s R_{j_i}}{\prod_{i=s+1}^4 R_{k_i}}$ and 
${\cal 
N}^2_{T,(r,s)}=\frac{V_{r+1}}{(2\pi)^{r+1}}2^{4-s}n^2$.} 
\end{center} 
\label{tb1}\end{table}
\smallskip

Finally, expanding to suitable order in windings and momenta the 
stability conditions for \dh-branes are found to match those given in 
Section~\ref{sec3}.

\section{Normalisations and partition functions of torsion charged branes}\label{appd}
\setcounter{equation}{0}

In this section we compute the one loop partition functions for the 
torsion-charged branes which couple to the twisted \NSNS 
sectors\footnote{For definiteness we consider the brane to be stretching 
along $x^5$ and $x^6$.} 
\ba
{\cal A}&=&\int_0^\infty\rmd 
l\bra{D(r,2)}e^{-lH_{\ms{c}}}
\ket{D(r,2)}\,,\\
{\cal M}_9&=&\int_0^\infty\rmd l\bra{C(5,4)}e^{
lH_{\ms{c}}}\ket{D(r,2)}+
\bra{D(r,2)}e^{-lH_{\ms{c}}}\ket{C(5,4)}\,,\\
{\cal M}_5&=&\int_0^\infty\rmd l\sum_{i=1}^{16}\bra{C(5,0),{\bf 
y}_i}e^{-lH_{\ms{c}}}
\ket{D(r,2)}+
\bra{D(r,2)}e^{-lH_{\ms{c}}}\ket{C(5,0),{\bf y}_i}\,,
\ea
and modular transform to the loop-channel, from which we may read of the 
values of $r$ and the normalisation of the boundary state for which the 
ground-state tachyon cancels. Evaluating the amplitudes\footnote{See equation~(\ref{torbrane})) for a 
definition of $\ket{D(r,2)}$.}
\ba
{\cal A}&=&\half{\cal N}_U^2\int_0^\infty\rmd l 
l^{(r-5)/2}\frac{f_3^8(q)-f_4^8(q)}{f_1^8(q)}
\prod_{i=5,6}\sum_{w_i\in\Z}e^{-\pi  
l(w_iR_i)^2}\prod_{j=7,8}\sum_{n_j\in\Z}e^{-\pi
l(n_j/R_j)^2}
\nonumber \\&&+\;\;2{\cal N}_T^2\int_0^\infty\rmd l l^{(r-5)/2}
\frac{f_3^4(q)f_2^4(q)}{f_1^4(q)f_4^4(q)}
\,, \\
{\cal M}_9&=&{\cal N}_U{\cal N}_{\ms{C}9}\int_0^\infty\rmd l
\frac{f_3^{r+1}(iq)f_4^{7-r}(iq)-f_4^{r+1}(iq)f_3^{7-r}(iq)}{f_1^{r+1}(iq)f_2^{7-r}(iq)2^{(r-7)/2}}
\prod_{i=5,6}\sum_{w_i\in 2\Z}e^{-\pi l(w_iR_i)^2}\,,\\
{\cal M}_5&=&16{\cal N}_U{\cal N}_{\ms{C}5}\int_0^\infty\rmd l
\frac{f_3^{r+1}(iq)f_4^{7-r}(iq)-f_4^{r+1}(iq)f_3^{7-r}(iq)}{f_1^{r+1}(iq)f_2^{7-r}(iq)2^{(r-7)/2}}
\prod_{j=7,8}\sum_{n_j\in 2\Z}e^{-\pi l(n_j/R_j)^2}\,,
\ea
and performing the modular transformation
\ba
{\cal A}&=&\frac{2^4{\cal N}_U^2R_7R_8}{R_5R_6}
\int_0^\infty\frac{\rmd t}{2t} 
(2t)^{-(r+1)/2}\frac{f_3^8(\tq)-f_2^8(\tq)}{f_1^8(\tq)}
\prod_{i=5,6}\sum_{n_i\in\Z}e^{-2\pi 
t(n_i/R_i)^2}\prod_{j=7,8}\sum_{w_j\in\Z}e^{-2\pi
t(w_jR_j)^2}
\nonumber \\&&+\;\;
2^4{\cal N}_T^2\int_0^\infty\frac{\rmd t}{2t} (2t)^{-(r+1)/2}
\frac{f_3^4(\tq)f_4^4(\tq)}{f_1^4(\tq)f_2^4(\tq)}\,, \\
{\cal M}_9&=&\frac{{\cal N}_U{\cal 
N}_{\ms{C}9}}{2R_5R_6}\int_0^\infty\frac{\rmd t}{2t}(2t)^{-(r+1)/2}
\nonumber \\&&\times\;\;
\frac{e^{i\frac{\pi}{4}(r-3)}f_4^{r+1}(i\tq)f_3^{7-r}(i\tq)-e^{-i\frac{\pi}{4}(r-3)}
f_3^{r+1}(i\tq)f_4^{7-r}(i\tq)}{f_1^{r+1}(i\tq)f_2^{7-r}(i\tq)2^{(r-7)/2}}
\prod_{i=5,6}\sum_{n_i\in\Z}e^{-2\pi t(n_i/R_i)^2}\,,\\
{\cal M}_5&=&8{\cal N}_U{\cal N}_{\ms{C}5}R_7R_8\int_0^\infty\frac{\rmd 
t}{2t}(2t)^{-(r+1)/2}
\nonumber \\&&\times\;\;
\frac{e^{i\frac{\pi}{4}(r-3)}f_4^{r+1}(i\tq)f_3^{7-r}(i\tq)-e^{-i\frac{\pi}{4}(r-3)}
f_3^{r+1}(i\tq)f_4^{7-r}(i\tq)}{f_1^{r+1}(i\tq)f_2^{7-r}(i\tq)2^{(r-7)/2}}
\prod_{j=7,8}\sum_{w_j\in\Z}e^{-2\pi t(w_jR_j)^2}\,.
\ea
With
\be
{\cal N}_U^2=\frac{V_{r+1}}{(2\pi)^{r+1}}n^2\frac{R_5R_6}{R_7R_8}\,,\qquad
{\cal N}_T^2=\frac{V_{r+1}}{(2\pi)^{r+1}}4n^2\,,
\ee
for some number $n$ the condition for the ground-state tachyon to cancel is
\be
32n^2=4\sqrt{2}n\sin(\frac{\pi(r-3)}{4})\,.
\ee
Since $r$ is an integer between $-1$ and $5$ one may easily verify that 
there are only two solutions (for positive ${\cal N}_U$)
\be
r=4\,,\spc n^2=\frac{1}{64}\qquad\mbox{or}\qquad r=5\,,\spc 
n^2=\frac{1}{32}\,.
\ee
Expanding to suitable order in momentum and winding one may confirm that 
the stability of these branes is as given in equation~(\ref{condtortor}).

A similar computation has been done for D-branes that couple only to the 
untwisted \NSNS sector, and one finds that for $r=4,5$ and $s=1,2,3$ the 
ground-state tachyon cancels for normalisations listed in Table~\ref{tab2} 
and the stability of these torsion branes is as given in~\ref{sec4}.
\begin{table}[hbt] \begin{center} \begin{tabular}{|c|c|c|c|}\hline
$(r,s)$ & (4,1), (4,3) & (4,2), (5,1), (5,3) & (5,2) \\ \hline &&& \\
$n^2$ & $\frac{1}{32}$ & $\frac{1}{16}$ & $\frac{1}{8}$ 
\\ &&&\\ \hline 
\end{tabular}
\caption{Normalisations of boundary states for $\Z_2$ torsion charged 
D-branes.
$n$ is related to the normalisations via ${\cal 
N}^2_{U,(r,s)}=\frac{V_{r+1}}{(2\pi)^{r+1}}n^2
\frac{\prod_{i=1}^s R_{j_i}}{\prod_{i=s+1}^4 R_{k_i}}$.}\label{tb2} 
\end{center} 
\label{tab2}
\end{table}
\smallskip

Note in particular that the $\Z_2$, $s=2$ torsion branes have the same 
normalisation 
as a boundary state corresponding to a bound states of two torsion branes 
from equation~(\ref{torbrane}) with opposite twisted \NSNS couplings.

\ed